\documentclass[aps,prl,reprint,superscriptaddress,showpacs,longbibliography,floatfix,footnoteinbib]{revtex4-1}
\usepackage[utf8]{inputenc}
\usepackage{color}
\usepackage{amsmath}
\usepackage{amssymb}
\usepackage{graphicx}
\usepackage[unicode=true,
 bookmarks=false,
 breaklinks=false,pdfborder={0 0 1},backref=false,colorlinks=true]
 {hyperref}
\hypersetup{pdftitle={Title},
 plainpages=false,pdfpagelabels,linkcolor=red,urlcolor=blue,citecolor=blue,pdfdisplaydoctitle=true,pdfduplex=DuplexFlipLongEdge,pdftex,breaklinks,bookmarks=false}

\makeatletter
\usepackage{units}\usepackage{wasysym}

\usepackage{bm}
\usepackage{braket}
\usepackage{mhchem}
\usepackage{isotope}

\makeatother

\begin{document}
\title{Effects of dilution in a 2D topological magnon insulator}
\author{Miguel S. Oliveira}
\affiliation{Centro de Física das Universidades do Minho e do Porto, LaPMET, Departamento
de Física e Astronomia, Faculdade de Ciências, Universidade do Porto,
4169-007 Porto, Portugal}
\author{T. V. C. Antão}
\affiliation{Laboratório de Instrumentação e Física Experimental de Partícuals
(LIP), University of Minho, 4710-057 Braga, Portugal}
\author{Eduardo V. Castro}
\affiliation{Centro de Física das Universidades do Minho e do Porto, LaPMET, Departamento
de Física e Astronomia, Faculdade de Ciências, Universidade do Porto,
4169-007 Porto, Portugal}
\affiliation{Beijing Computational Science Research Center, Beijing 100084, China}
\author{Nuno Peres}
\affiliation{Centro de Física das Universidades do Minho e do Porto (CF-UM-UP)
e Departamento de Física, Universidade do Minho, P-4710-057 Braga,
Portugal}
\affiliation{International Iberian Nanotechnology Laboratory (INL), Av Mestre José
Veiga, 4715-330 Braga, Portugal}
\affiliation{POLIMA---Center for Polariton-driven Light-Matter Interactions, University
of Southern Denmark, Campusvej 55, DK-5230 Odense M, Denmark}
\begin{abstract}
We study the effect of diluting a two-dimensional ferromagnetic insulator
hosting a topological phase in the clean limit. By considering the
ferromagnetic Heisenberg model in the honeycomb lattice with second
nearest-neighbor Dzyanshikii-Moriya interaction, and working in the
linear spin-wave approximation, we establish the topological phase
diagram as a function of the fraction $p$ of diluted magnetic atoms.
The topological phase with Chern number $C=1$ is robust up to a moderate
dilution $p_{1}^{*}$, while above a higher dilution $p_{2}^{*}>p_{1}^{*}$
the system becomes trivial. Interestingly, both $p_{1}^{*}$ and $p_{2}^{*}$
are below the classical percolation threshold $p_{c}$ for the honeycomb
lattice, which gives physical significance to the obtained phases.
In the topological phase for $p<p_{1}^{*}$, the magnon spectrum is
gapless but the states filling the topological, clean-limit gap region
are spatially localized. For energies above and below the region of
localized states, there are windows composed of extended states. This
is at odds with standard Chern insulators, where extended states occur
only at single energies. For dilutions $p_{1}^{*}<p<p_{2}^{*}$, the
two regions of extended states merge and a continuum of delocalized
states appears around the middle of the magnon spectrum. For this
range of dilutions the Chern number seems to be ill defined in the
thermodynamic limit, and only for $p>p_{2}^{*}$, when all states
become localized, the system shows $C=0$ as expected for a trivial
phase. Replacing magnetic with non-magnetic atoms in a systems hosting
a magnon Chern insulator in the clean limit puts all the three phases
within experimental reach.
\end{abstract}
\maketitle

\section{Introduction}

Electronics based on the spin degree of freedom (Spintronics) harbor
well-known advantages from the present silicon-based technologies~\citep{Rajput2022}.
Magnonics~\citep{Lenk2011}, a subfield of Spintronics, addresses
the use of spin waves (magnons) to transmit and process information~\citep{Serga2004,Demidov2009,Jorzick2002,Podbielski2006}.
By replacing charge currents by spin currents, it is possible to avoid
Joule heating, making these technologies of great practical interest~\citep{Barman2021}.
These waves have already been measured in Yttrium Iron Garnet thin-films
(YIG)~\citep{Jungfleisch2015}. On the other hand, the study of magnons
furnishes fundamental insight about collective excitations and low-energy
properties of quantum magnetic systems.

The observation of the anomalous thermal Hall Effect for magnons --
intrinsic transverse heat transport response to a longitudinal temperature
gradient -- in the insulating ferromagnet \ce{Lu_2V_2O_7} with a
pyrochlore structure~\citep{Onose2010,PMID:25838381}, and more recently
in a Kagomé magnet~\citep{Hirschberger2015}, confirmed that magnonic
systems may host topological properties~\citep{Chisnell2015} just
like other systems with bosonic quasiparticles, such as photons~\citep{Rikken1996}
and phonons~\citep{Zhang2010}. Topology was first introduced to
condensed matter systems through the 2D electron gas in the quantum
Hall regime~\citep{TKNN82} and through the electronic models of
Haldane~\citep{PhysRevLett.61.2015} and Kane-Mele~\citep{KM05}
in the honeycomb lattice. Soon after, topological insulators, a novel
type of materials characterized by an insulating bulk and metallic
boundary states, became a hot spot of research for their exotic properties
and possible applications~\citep{Fu2010,Vijay2015,Dennis2002}. The
presence of edge states in these materials is ensured by the bulk-edge
correspondence, a topological property which dictates that edge states
properties are deeply connected to the bulk and its symmetries. For
small disorder which does not break essential symmetries, these edge
states are immune to back-scattering making them the ideal transport
states~\citep{Lado2015,Murakami2011}. In the case of neutral quasi-particles
such as magnons, the non trivial topology may come from the spin-orbit
interaction, imposed by Dzyaloshinskii-Moriya (DM)~\citep{Moriya1960,McClarty2021}
contribution to the exchange interaction between localized magnetic
moments.

A key property of topological insulators is their robustness to moderate
disorder~\citep{Wu2017,Li2019,Xiao2010}, making them attractive
for real applications where disorder cannot be avoided. Disorder may
even be an essential ingredient for the experimental observation of
a topological response, as is the case of the quantum Hall effect~\citep{Kramer1993}.
In some cases, disorder may also benefit the appearance of topological
properties in regimes where the unperturbed model is trivial. The
topological Anderson insulator is a well known example of this class~\citep{Groth2009,Li2009,Goncalves2018},
which has recently been realized in cold atoms~\citep{Meier2018},
photonic crystals~\citep{Stutzer2018}, and electrical circuits~\citep{Zhang2019}.
Disorder induced nontrivial topology has also been proposed~\citep{Li2020,yang2021higher,Agarwala2020,wang2021structural,Peng2022,Loio2023}
and observed~\citep{Zhang2020} in higher-order topological insulators.
For topological magnon insulators, however, the role of disorder has
been much less appreciated.

A particularly interesting type of disorder in magnetic systems is
dilution, which can be achieved by replacing a certain fraction $p$
of magnetic atoms by non-magnetic ones. Diluted quantum magnets have
been studied previously, both diluted ferromagnets~\citep{Edwards1971}
and antiferromagnets~\citep{Wan1991}. Since long range order is
limited by the classical percolation threshold of the underlying lattice,
the question then was whether a quantum phase transition could be
induced by dilution prior to the expected classical percolation transition.
This question is even more relevant in 2D due to the low dimensionality,
but a stochastic series expansion approach for the diluted 2D antiferromagnetic
Heisenberg model clearly showed that the percolating cluster at $p_{c}$
(the critical classical dilution threshold) showcases long-range order~\citep{Sandvik2001,Sandvik2002}.
It was later shown that critical exponents involving dynamical correlations
are different from the classical percolation values, even though the
transition is driven by the underlying classical percolation~\citep{Vojta2005}.

In this article we study the interplay between nontrivial topology
and magnetic dilution in 2D topological magnon insulators. We address
in particular the question of whether a topological transition takes
place with increasing dilution and whether or not there is a relation
with the classical percolation threshold of the underlying lattice.
Since topological properties are deeply related with localization
and transport properties, we also study the effect of dilution on
localization of the magnon states. Moreover, dilution falls in the
strong disorder classification even for small percentages of vacancies.
There is no consensual answer to whether localization properties for
this type of disorder follow a one parameter scaling theory~\citep{Abrahams1979},
like other types of disorder such as Anderson disorder~\citep{Anderson1958}.
Using the analysis of the transmission coefficient associated to a
simple nearest neighbor model in a diluted lattice, it was shown that
for increasing disorder and for a fixed finite energy, the system
undergoes a transition from delocalized states to power-law localized
states, and for large dilution the system is exponentially localized~\citep{Islam2008,Dillon_2014}.
We anticipate that the complexity of the localization properties for
this type of disorder may introduce interesting topological behaviors.

The paper is organized as follows: In Sec.~\ref{sec:Model-and-Methods},
we introduce the model and the methods used to characterize the system.
The topological, spectral, and localization properties are discussed
in Sec.~\ref{sec:Results}. In Sec.~\ref{sec:Discussion} we provide
a thorough discussion of the obtained results. The key results are
summarized in Sec.~\ref{sec:Conclusions} and some conclusions are
drawn. In Appendix~\ref{secap:Classical-percolation} we compute
the classical site percolation threshold $p_{c}$ for the diluted
honeycomb lattice with first and second neighbor connections.

\section{Model and Methods}

\label{sec:Model-and-Methods}

\subsection{Magnons Model}

\subsubsection{Clean case}

We consider the Heisenberg model for localized spins with DM~\citep{Moriya1960}
and Zeeman exchange interaction for a ferromagnetic system in the
honeycomb lattice with the Hamiltonian written as

\begin{equation}
{\cal H}=J\sum_{\left\langle ij\right\rangle }{\bf S}_{i}\cdot{\bf S}_{j}+\sum_{\left\langle \left\langle ij\right\rangle \right\rangle }{\bf D}_{ij}\cdot\left[{\bf S}_{i}\times{\bf S}_{j}\right]-B\sum_{i}S_{i}^{z},\label{eq:H_DM}
\end{equation}
where $\left\langle \dots\right\rangle $ and $\left\langle \left\langle \dots\right\rangle \right\rangle $
refers respectively to the summation over the NN and NNN elements
of the lattice and the coupling ${\bf D}_{ij}$ in the honeycomb lattice
is given by $\nu_{ij}D\hat{z}$, with $\nu_{ij}=2/\sqrt{3}\left({\bf d}_{1}\times{\bf d}_{2}\right)\cdot{\bf z}=\pm1$,
where ${\bf d}_{1},{\bf d}_{2}$ are two unit vectors along the bonds
connecting the NNN $\left\langle \left\langle i,j\right\rangle \right\rangle $.
In the following analysis we fix $\hbar=1$. Considering the low temperature
regime, it is possible to study this model by focusing on deviations
from the ferromagnetic ground state, and using Holstein-Primakoff
formalism~\citep{Holstein1940} to define the creation $(a_{i}^{\dagger})$
and annihilation $(a_{i})$ bosonic operators associated to spin deviations
$a_{i}^{\dagger}a_{i}=n_{i}=S-S_{i}^{z}$, where $n_{i}$ is the number
operator of the site $i$. The spin operators may then be rewritten
as

\begin{align}
S_{i}^{-} & =\sqrt{2S}a_{i}^{\dagger}\sqrt{1-\frac{a_{i}^{\dagger}a_{i}}{2S}}\approx\sqrt{2S}a_{i}^{\dagger}\nonumber \\
S_{i}^{+} & =\sqrt{2S}\sqrt{1-\frac{a_{i}^{\dagger}a_{i}}{2S}}a_{i}\approx\sqrt{2S}a_{i},\label{eq:HP}
\end{align}
where we only kept the zeroth order term, disregarding interactions
between bosons, since we will focus only in the low energy regime
where $a_{i}^{\dagger}a_{i}\ll2S$, the linear spin-wave approximation.
Bosonic statistics do not have a limit on the occupation factor, however,
in this approximation the maximum occupation number cannot exceed
the number of deviations needed to surpass the minimum spin value.

We can rewrite the Hamiltonian with respect to bosonic operators as

\begin{align}
{\cal H}= & 2JS\sum_{\left\langle ij\right\rangle }\left(a_{i}^{\dagger}a_{i}-a_{i}^{\dagger}a_{j}\right)\nonumber \\
 & +2iSD\sum_{\left\langle \left\langle ij\right\rangle \right\rangle }\nu_{ij}a_{i}^{\dagger}a_{j}-B\sum_{i}a_{i}^{\dagger}a_{i},\label{eq:H_DM_bosonic}
\end{align}
where we have omitted the constant term associated to the ferromagnetic
ground state energy. Making use of translational invariance to write
${\cal H}$ in the momentum basis, it is possible to show that the
spectrum of the model follows the dispersion relation

\begin{equation}
E({\bf k})=2JS\left(3\pm\sqrt{\left|f({\bf k})\right|^{2}+m({\bf k})^{2}}\right),\label{eq:Haldane_k}
\end{equation}
where

\begin{align}
m({\bf k}) & =-2\frac{D}{J}\left(\sin({\bf k}\cdot{\bf b}_{1})-\sin\left({\bf k}\cdot({\bf b}_{1}-{\bf b}_{2})\right)-\sin({\bf k}\cdot{\bf b}_{2})\right)\nonumber \\
f({\bf k}) & =1+\exp(i{\bf b}_{1}\cdot{\bf k})+\exp(i{\bf b}_{2}\cdot{\bf k}),\label{eq:mkfk}
\end{align}
${\bf b}_{1},{\bf b}_{2}$ being the reciprocal lattice vectors.

\subsubsection{Non-trivial topology in the clean limit}

It becomes clear in Eq.~\eqref{eq:H_DM_bosonic} that in the lowest
order spin-wave approximation the system is described by a single-particle
tight binding model. Apart from the absence of a trivial mass, this
bosonic model is equivalent to the ferminonic system introduced by
Haldane~\citep{PhysRevLett.61.2015}, hosting topological phases
characterized by the topological index Chern number $C$. Given an
eigenstate $|\Psi_{{\bf k}}^{n}\rangle$ of our system associated
to the band $n$, the Berry curvature may be defined as
\begin{equation}
\Omega_{xy}^{(n)}({\bf k})=-2\text{Im}\left(\left\langle \partial_{k_{x}}\Psi_{{\bf k}}^{n}\Big|\partial_{k_{y}}\Psi_{{\bf k}}^{n}\right\rangle \right).\label{eq:Berry_curvature}
\end{equation}
The Chern number corresponds to the integral of Berry curvature in
Eq.~\eqref{eq:Berry_curvature} over the first Brilloin zone,

\begin{align}
C_{n} & =\frac{1}{2\pi}\iint_{1ZB}\Omega_{xy}^{n}d^{2}{\bf k}.\label{eq:Chern_number}
\end{align}
In the trivial phase, both bands have a null Chern number while in
the topological phase the valence and conduction band have $C=\pm1$,
keeping the overall Chern number null.

Even though this model is a bosonic model and the notion of band filling
does not apply like in the equivalent fermionic model, the topological
properties of a non-interacting system depend exclusively on the single
particle Hamiltonian, which is independent of the statistical descriptions
of the particles which inhabit the model. In the bosonic case, the
transport properties at low temperature will not be described by the
states at the middle of the spectrum, but one can always force an
excitation to have a certain energy with a external stimulus. For
example, it is possible to excite the edge states in a magnon model,
provided that the system is in a topological phase~\citep{Shindou2013,Zhang2013b,Mook2014}.

\subsubsection{Diluted Magnons}

Diluted lattices are simulated by assuming that each site is independently
and randomly occupied with a localized spin with probability $1-p$
and it is empty with probability $p$. Whenever a site is assigned
as a vacancy, it is removed from the tight binding basis. A \emph{cluster
}is defined as a set of occupied neighboring sites, which in the terminology
of graph theory represents a connected graph. In the spirit of the
tight binding approach, we define the neighborhood (set of edges of
one vertex) of a site as the set of occupied sites which share non-zero
hopping terms. In every disorder configuration we choose random occupied
sites and in the end retain solely the largest connected cluster since
this will be the one with physical relevance. After collecting the
cluster, we build the Hamiltonian of the system with the basis of
atomic orbitals of occupied sites only belonging to the largest connected
cluster.

\begin{figure}
\begin{centering}
\includegraphics[width=0.95\columnwidth]{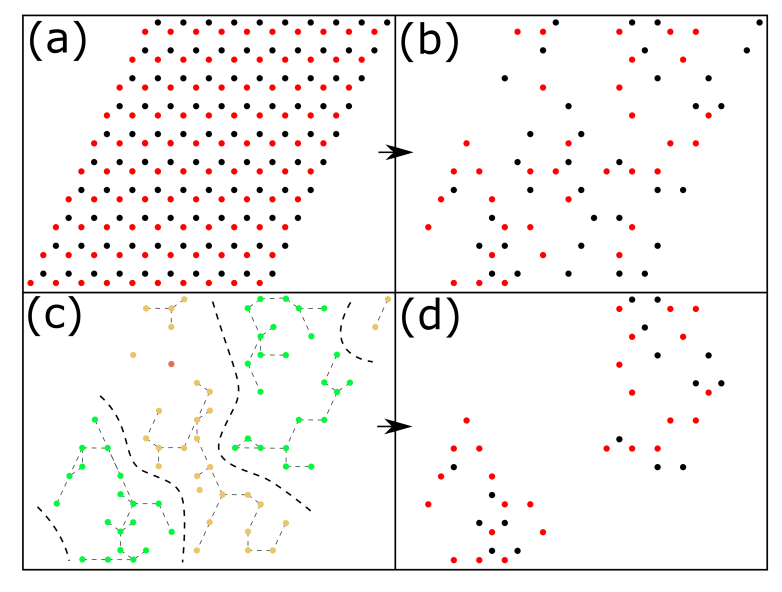}
\par\end{centering}
\caption{\label{fig:Obtaining-the-LC}To obtain the largest cluster of a diluted
lattice, we start from a clean honeycomb lattice (a), where the red
and black colors make the bipartite nature of the lattice explicit.
Running through all the sites in the lattice, we randomly assign the
site as vacant with probability $p$ and in this case we remove it
from our graph (b). After choosing all the vacant sites, we collect
the different clusters of disconnected sites, as shown in panel (c)
with different colors. The model we will consider has next-to-nearest
neighbors hoppings, hence we consider that for 2 sites to belong to
the same cluster they should be either first neighbors or second neighbors,
considering periodic boundary conditions. From all the clusters, we
retain just the one with more sites and encode it in a tight binding
Hamiltonian (d).}
\end{figure}

Periodic boundary conditions must be employed in order to calculate
bulk properties. This modification of the lattice implies occupied
sites in two distinct boundaries may still be connected and belong
to the same cluster, as exemplified in Fig.~\ref{fig:Obtaining-the-LC}.
A complete finite size honeycomb lattice is shown in Fig.~\ref{fig:Obtaining-the-LC}(a).
In Fig.~\ref{fig:Obtaining-the-LC}(b) we assign randomly the vacant
sites, and in Fig.~\ref{fig:Obtaining-the-LC}(c) we cluster the
disjoint sets of sites. By keeping only the largest set of sites we
obtain the largest connected cluster, shown in Fig.~\ref{fig:Obtaining-the-LC}(d).

As one increases the percentage of vacancies in the simulated lattice,
for percentages larger than a critical value, the size of the largest
cluster no longer scales with the same dimensionality as the size
of the clean system. It is not possible to take the thermodynamic
limit for $p>p_{c}$, as the simulated system has no dependence on
the original size of the clean system. This critical value is the
so-called classical percolation threshold $p_{c}$ and only depends
on the lattice geometry~\citep{Suding1999,Ziff1992}. A second and
equivalent definition of this quantity, which highlights the physical
relevance of this property, is the following: the probability of having
a macroscopic path composed of connected sites on a simulated diluted
lattice is $0$ if $p>p_{c}$ and is $1$ if $p<p_{c}$. This implies
that there is no transport nor long range order for $p>p_{c}$. Although
the classical percolation threshold for the honeycomb lattice with
nearest neighbors connections is well known to be $p_{c}\approx0.302957$~\citep{Suding1999},
there are no results in the literature for the honeycomb with first
and second nearest neighbors connections. We determined this value
to be $p_{c}=0.640\pm0.005$ (see Appendix~\ref{secap:Classical-percolation}).

In the diluted regime, the Hamiltonian is rewritten as

\begin{align}
{\cal H}= & 2JS\sum_{\left\langle ij\right\rangle }\eta_{i}\eta_{j}\left(a_{i}^{\dagger}a_{i}-a_{i}^{\dagger}a_{j}\right)\nonumber \\
 & +2iSD\sum_{\left\langle \left\langle ij\right\rangle \right\rangle }\eta_{i}\eta_{j}\nu_{ij}a_{i}^{\dagger}a_{j}-B\sum_{i}\eta_{i}a_{i}^{\dagger}a_{i},\label{eq:Diluted_Magnons}
\end{align}
where dilution is introduced in the Hamiltonian with the random variables
$\eta_{i}$, which take the values $0$ or $1$ depending whether
the site $i$ exists in the lattice being simulated or not. Notice
that the first term of the Hamiltonian in Eq.~\eqref{eq:Diluted_Magnons}
introduces a dependency on the number of occupied neighbors of one
site to the respective on-site energy, which is not present in the
Haldane model. In this work we fixed $2JS\equiv t$, $2SD=0.1t$,
$B=0$, and use $t$ as the energy unit.

Disorder breaks translational invariance, meaning that the Hamiltonian
eigenstates will no longer be Bloch states with well defined Bloch
momentum and band index. Nevertheless, it is still possible to extend
the formalism applied in undiluted systems, and compute the Chern
number with a real space approach, as discussed next.

\subsection{Methods}

\subsubsection{Chern number}

The computation of the Chern number can be extended to systems with
broken translational invariance employing a super-cell approach with
twisted boundary conditions in real space. With Fukui’s method~\citep{Fukui2005},
one can replace the continuous integral in Eq.~\ref{eq:Chern_number}
by a discrete sum over boundary twists $\boldsymbol{\theta}$. Additionally,
with the coupling matrix method of Ref.~\citep{Zhang2013}, we can
reduce the sum to twist angles associated to periodic boundary conditions.
This method requires exact diagonalization of two matrices with size
proportional to the system's size: the Hamiltonian, and the coupling
matrix~\citep{Zhang2013}.

To minimize finite size effects, we average the Chern number over
disorder realizations. Even though the Chern number is strictly quantized
for each disorder configuration, the averaged Chern number may not
be. A finite size scaling analysis is performed when needed in order
to infer the thermodynamic limit behavior.

\subsubsection{Density of states}

In order to build the gapless-gapped picture for the model, we used
the Kernel Polynomials Method (KPM)~\citep{Weisse2006} to compute
the density of states through a Chebyshev polynomials series expansion,
equipped by the Jackson's Kernel. The linear scaling of this method
with the system size allowed us to explore systems of the order of
$N=1024^{2}$ unit cells.

A finite cutoff of the series expansion introduces an artificial smoothing
in abrupt changing parts of the spectrum, which may lead to wrong
interpretations about the existence of a gap for small enough gaps.
In these cases, an analysis of the convergence of the DOS in the gap
with respect to the cutoff order of the series expansion was employed
in hopes of minimizing the uncertainty of the limits of the gapped-gapless
phases.

\subsubsection{Localization}

\label{subsec:Localization}

The localization behavior of the system was explored using the standard
tools: level spacing statistics~\citep{Evers2008} to try to observe
a transition from a GUE (Gaussian Unitary Ensemble) distribution to
a Poisson distribution as states become localized, and Transfer Matrix
Method (TMM)~\citep{Kramer1993} to compute the scaling of the correlation/localization
length across the spectrum for different dilution percentages.

In order to identify the probability distribution of the level spacings,
we looked at the distribution of the ratio between consecutive level
spacings, $r_{n}=\min(s_{n},s_{n+1})/\max(s_{n},s_{n+1})$, where
$s_{n}=\epsilon_{n+1}-\epsilon_{n}$ are the level spacings. This
quantity is known to have better behavior in regions of the spectrum
with low density of states~\citep{Oganesyan2007,Atas2013}.

In order to implement the TMM, we had to assign a high on-site energy
$V_{\text{vacancy}}$ to the vacant sites, plus a very small hopping
term $\delta t$ from adjacent sites to these vacant sites, instead
of removing the vacant sites from our tight binding basis. We used
$V_{\text{vacancy}}=1000t$ and $\delta t/t=1/1000$. The limit $V_{\text{vacancy}}\to\infty,\delta t\to0$,
where both implementations of dilution should be equivalent, is discussed
in Sec.~\ref{sec:Discussion}.

\section{Results}

\label{sec:Results}

\subsection{Topological phase diagram}

\begin{figure}
\begin{centering}
\includegraphics[width=0.95\columnwidth]{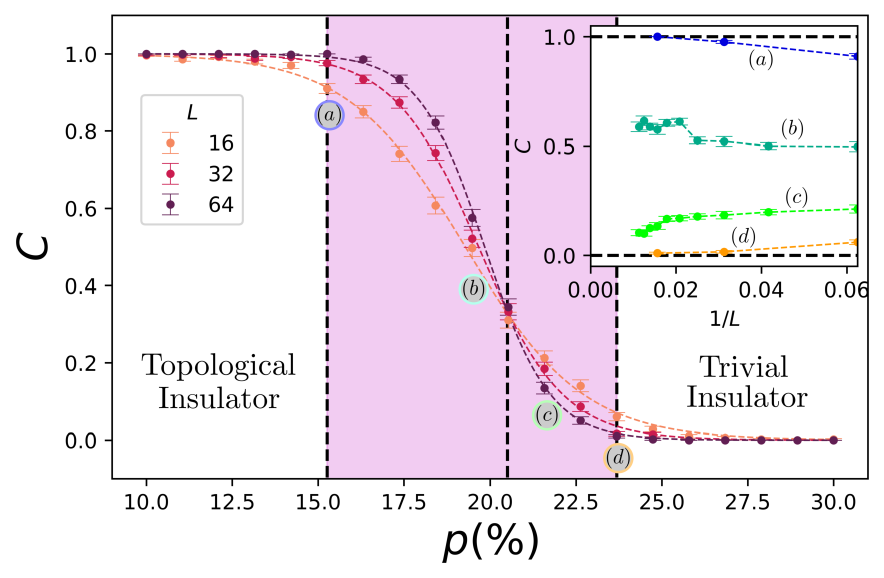}
\par\end{centering}
\caption{Averaged Chern number for different system sizes ($L$ is the total
number of unit cells in one spacial direction) and for different percentages
of vacancies. The standard deviation errors obtained by averaging
over $500$ disorder realizations are shown for each data point. Inset:
finite size scaling analysis for the averaged Chern number at specific
dilution values: $p=15.3\%$~(a), $p=19.5\%$~(b), $p=21.6\%$~(c),
and $p=23.7\%$~(d).\label{fig:Computed-Chern-number}}
\end{figure}

As can be seen in Fig.~\ref{fig:Computed-Chern-number}, the topological
phase is robust up to relatively high percentages of vacancies, persisting
with a well defined Chern number $C=1$ up to $p\approx15\%$. In
the inset of Fig.~\ref{fig:Computed-Chern-number} it is clearly
seen that for $p=15.3\%$, labeled with~(a) in the inset, we have
an averaged Chern number $C\rightarrow1$ as $1/L\to0$. For large
enough dilution, which we overestimate as $p\gtrsim23\%$, the system
enters the trivial phase. In this region of the phase diagram $C\rightarrow0$
as $1/L\rightarrow0$, as shown for $p=23.7\%,$labeled with~(d)
in the inset of Fig.~\ref{fig:Computed-Chern-number}. The system
becomes a trivial insulator before reaching the classical percolation
threshold for the honeycomb lattice, $p_{c}=30\%$, and long before
the threshold for the honeycomb lattice with first and second nearest
neighbors, $p_{c}=64\%$, rendering the trivial phase physically achievable.

Between the topological and the trivial phases shown in Fig.~\ref{fig:Computed-Chern-number},
there is a highlighted region for $15\%\lesssim p\lesssim23\%$ where
the averaged Chern number crosses over from $C=1$ to $C=0$ as $p$
increases. Our numerical analysis is not conclusive on whether this
region shrinks to a single critical dilution in the thermodynamic
limit or remains finite. Nevertheless, a finite size scaling analysis
seems to indicate that the averaged Chern number does not converge
to quantized values as we approach the thermodynamic limit. This is
exemplified in the inset of Fig.~\ref{fig:Computed-Chern-number}
for $p$ values labeled with~(b) and~(c). The presence of a crossover
region which does not shrink to a single critical dilution is also
justified by the localization behavior of the system to be discussed
in Sec.~\ref{subsec:Localization-properties}.

\subsection{Spectral and localization properties}

\subsubsection{Spectral properties}

\begin{figure}
\begin{centering}
\includegraphics[width=0.95\columnwidth]{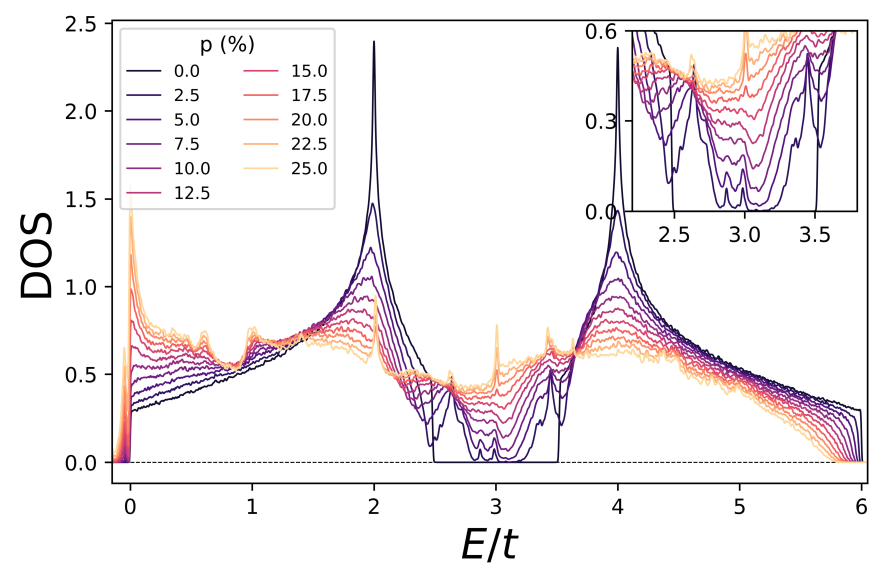}
\par\end{centering}
\caption{DOS for different percentages of dilution, including both the topological
and the trivial phases, as well as the crossover region. We used systems
with size $1024\times1024$ and included $1024$ polynomials in the
KPM.\label{fig:DOS-for-different-p}}
\end{figure}

In Fig.~\ref{fig:DOS-for-different-p} we show the magnon DOS for
different values of dilution $p$ obtained using KPM. It is obvious
that even for a small percentage of vacancies (above $p\simeq5\%$)
the gap is completely filled. This implies that the topological phase
becomes gapless for small values of dilution. The gap remains closed
as disorder increases, even when the system becomes trivial.

The diagonal disorder seen in Eq.~\eqref{eq:Diluted_Magnons} destroys
the particle-hole symmetry of the spectrum for finite dilution, something
that does not appear in analogous fermionic systems, like the Haldane
model. This asymmetry shifts the middle of the spectrum to lower energies
as dilution increases.

\subsubsection{Localization properties}

\label{subsec:Localization-properties}

In order to provide a more complete analysis of the topological phase
transition, we have also characterized the localization properties
of the system for different values of $p$. The ratio between consecutive
level spacings $r_{n}$, introduced in Sec.~\ref{subsec:Localization},
allows to distinguish between localized and extended states. For localized
states $r_{n}$ should follow the Poisson distribution, while for
extended states in systems which break time reversal symmetry we expect
the Gaussian Unitary Ensemble (GUE) distribution~\citep{Evers2008}.

\begin{figure}[htp]
\begin{centering}
\includegraphics[width=0.98\columnwidth]{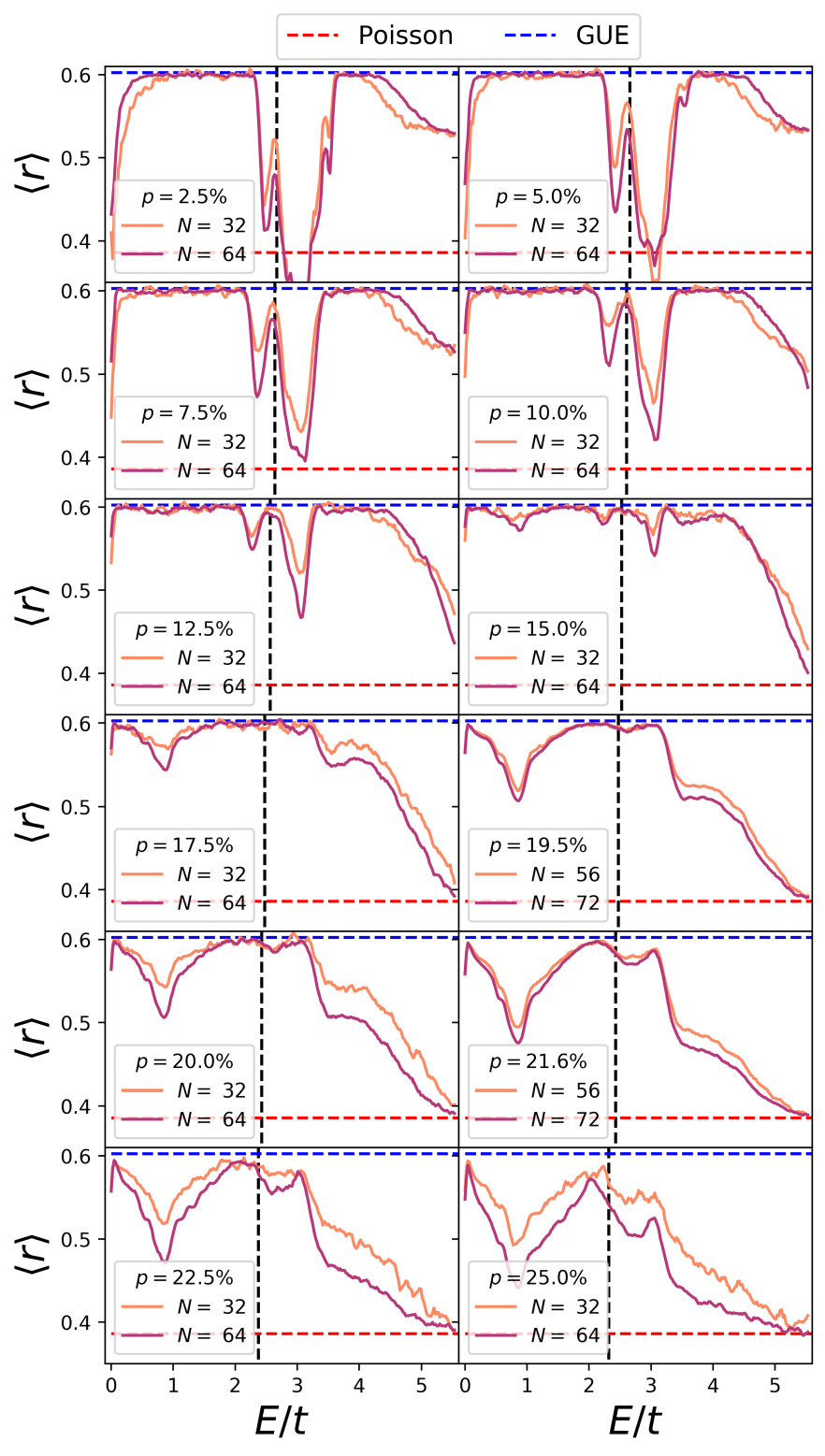}
\par\end{centering}
\centering{}\caption{Finite size scaling for the average value $\left\langle r\right\rangle $
of the distribution of the ratio between consecutive energy spacings
as a function of energy for increasing values of dilution percentages
$p$. We compare this average with the one expected for the GUE (horizontal
dashed blue line) and Poisson distribution (horizontal dashed red
line). The vertical dashed black line marks the middle of the spectrum,
marking also the upper limit of integration of the Berry curvature.
The statistics are averaged over a total number of $2048$ realizations
of disorder for $N=32^{2}$ and $256$ realizations of disorder for
$N=64^{2}$, and the bin's size used for the statistics is of the
order of $0.05t$.\label{fig:Average-value-of-ratio}}
\end{figure}

In Fig.~\ref{fig:Average-value-of-ratio} we show the first moment
of the distribution, $\left\langle r\right\rangle $, indicating as
dashed horizontal lines the expected values for the Poisson and GUE
distributions. With increasing dilution up to $p\lesssim15\%$, there
are two persistent regions, one below the middle of the spectrum (signaled
by the vertical dashed line) and the other above it, where $\left\langle r\right\rangle $
seems to converge to GUE with increasing system size. These two regions
are separated by a region of localized states around the middle of
the spectrum, where $\left\langle r\right\rangle $ is converging
to the Poisson value. The presence of two finite energy regions where
states are delocalized is at odds with the behavior of quantum Hall
systems~\citep{Kramer1993,wang2014anti} and conventional Chern insulators~\citep{Onoda2003,Onoda2007,Castro2016,Goncalves2018},
where extended states appear only at isolated single energies. However,
site dilution (or vacancies) is a particular type of disorder where
unconventional behavior is to be expected~\citep{Islam2008,Dillon_2014}.
Yet another important remark to be made is the presence of localized
states around the middle of the spectrum, separating the two regions
of extended states. Since the spectrum is gapless in the topological
phase, it is the feature of having localized states filling the gap
which allows the system to retain the topological properties reminiscent
from the clean system.

If we further increase the percentage of vacancies, it can be seen
in Fig.~\ref{fig:Average-value-of-ratio} that for $15\%\lesssim p\lesssim23\%$
the states previously localized for smaller $p$ values become extended,
forming a continuum of extended states around the middle of the spectrum.
We believe this continuum to be responsible for the crossover behavior
observed in Fig.~\ref{fig:Computed-Chern-number} regarding the topological
character of the system. The presence of this continuum prevents the
usual ``levitation and pair annihilation'' of the extended states
carrying the topological index~\citep{Laughlin1984,Onoda2007,Prodan2011}
and the existence of critical dilution at which the topological transition
would take place. As soon as dilution reaches the trivial phase, all
the spectrum becomes localized as is possible to observe in Fig.~\ref{fig:Computed-Chern-number}
for $p=25\%$.

\begin{figure}[htp]
\begin{centering}
\includegraphics[width=0.98\columnwidth]{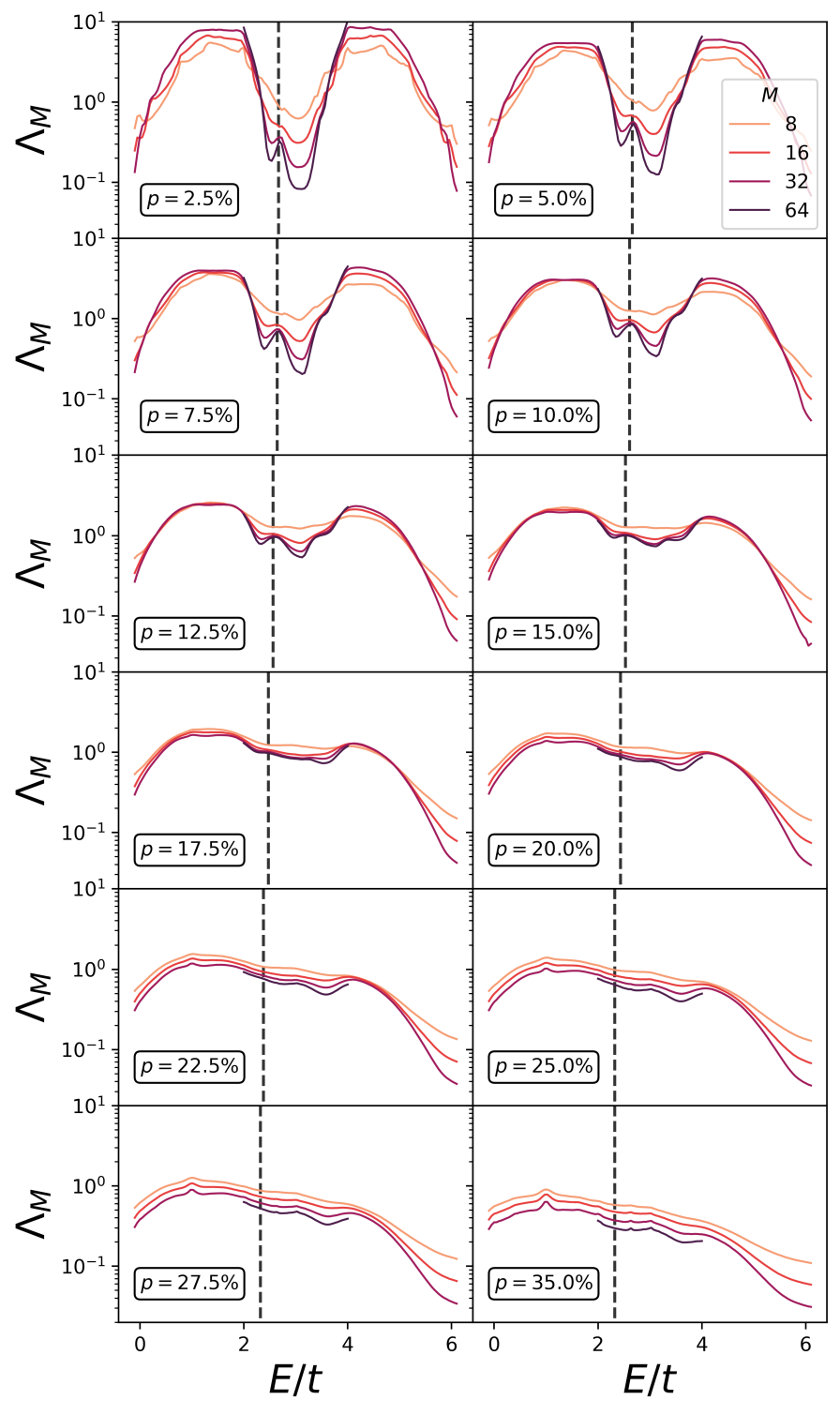}
\par\end{centering}
\centering{}\caption{Finite size scaling analysis of the normalized localization length
$\Lambda_{M}$ with the transverse size of the ribbon $M$ in the
TMM for different concentrations of vacancies. The longitudinal length
used ensures that the estimated error had converged to under $1\%$.
The energy resolution is of the order of $0.05t$. \label{fig:Loc_lenght}}
\end{figure}

In order to enrich our conclusions regarding the localization properties,
we computed the localization length with the TMM. As can be seen in
Fig~\ref{fig:Loc_lenght}, for small dilution up to $p\lesssim15\%$
we observe two regions in which the normalized localization length
$\Lambda_{M}$ is scale invariant, corresponding to delocalized states.
These two regions are separated by a region where $\Lambda_{M}$ decreases
with $M$, signaling the presence of localized states. This agrees
well with the level spacing statistics results of Fig.~\ref{fig:Average-value-of-ratio}.
In the crossover region $15\%\lesssim p\lesssim23\%$, we also see
the presence of a continuum of extended states spanning a large region
around the middle the spectrum (note the collapse of $\Lambda_{M}$
for the larger $M$ values). This confirms the unconventional localization
properties of this system when compared with conventional disordered
Chern insulators~\citep{Goncalves2018}. Upon reaching the trivial
phase, all the states become localized, as is possible to observe
for $p\gtrsim23\%$, where $\Lambda_{M}$ decreases with the transverse
size of the system, $M$, for the entire spectrum. Even though the
TMM results and those from level spacing statistics are in qualitative
agreement, there are noticeable quantitative differences which we
discuss in Sec.~\ref{sec:Discussion}.

\subsubsection{Spin-wave stability}

As dilution increases, a finite density of states with negative energies
appears very close to zero, as can be clearly seen in Fig.~\ref{fig:DOS-for-different-p}.
Negative energy eigenstates would imply that the ferromagnetic state
is not the true ground state of the system, rendering the usage of
Holstein-Primakoff transformation not valid. This problem can be easily
surpassed by making use of a finite magnetic field, which would make
a rigid shift of the spectrum. This shift would open a energy gap
between the ferromagnetic state and the lowest energy excitations,
thus stabilizing the spin-waves. The presence of single-ion magnetic
anisotropy would also give rise to a term in the Hamiltonian with
similar effects.

\section{Discussion}

\label{sec:Discussion}

The unconventional localization properties in the crossover region
support the conjecture that a topological index cannot be defined
for this range of dilutions since the system has delocalized magnons
in a large window around the middle of the spectrum (equivalent to
a metallic phase in fermionic systems). This is in agreement with
the results of Fig~\ref{fig:Computed-Chern-number} for the averaged
Chern number in the crossover region, which does not seem to converge
to a quantized value. The topological to trivial phase transition
occurs when the localized states, which separate the energy windows
with extended states, become localized. As dilution is increased,
this enables a continuous flow of Berry curvature between the two
regions with delocalized states. We speculate that the presence of
a smooth crossover transition occurs because of the appearance of
this continuum near the transition.

In fermionic models belonging to the same symmetry class similar behavior
was already observed: the transition from the topological phase to
the trivial one on increasing disorder occurs through a metallic phase~\citep{avishai2015criticalMetal,Qiao2016,avishai2016Metal,Yang2021},
for a finite range of disorder values and not at a single critical
disorder. The key point to obtain the metallic behavior in these works
is the presence of spin-flip hopping disorder. Even though a clear
relation between this latter type of disorder and dilution is hard
to establish, we notice that, in the clean limit, the two sublattices
of the honeycomb lattice ensure a two-component wavefunction similar
to spin-$1/2$ systems; around the middle of the spectrum it is even
possible to define a pseudo-spin from the sublattice degree of freedom~\citep{NGPrmp}.
Pushing forward the connection, nearest neighbor hopping disorder
would mix the two components of the wave function in a similar way
as spin-flip processes mix spin-up and spin-down. Finally, dilution
may be regarded as an extreme case of hopping disorder, where certain
random hoppings are set to zero.

Another point worth mentioning is the reason for some quantitative
differences between the results from the TMM in Fig.~\ref{fig:Loc_lenght}
and the level spacing statistics in Fig.~\ref{fig:Average-value-of-ratio}.
In the level spacing statistics we make use of exact diagonalization
of the Hamiltonian associated solely to the largest cluster for a
given configuration of vacancies, while for the TMM we cannot remove
the small isolated clusters from the computation. This means that
this second method opens the possibility of hosting for the same energy
value, both localized states in the independent clusters and localized
states over the physical cluster, which hybridize through the small
hopping used to simulate vacancies (see Sec.~\ref{subsec:Localization}).
We believe this to be the reason for the absence of scaling of $\Lambda_{M}$
with $M$ in sections of the spectrum for which the level spacing
statistics seems to tell us unambiguously that we are in the presence
of localized states (example around $E=4t$ for dilutions in the crossover
region).

\section{Conclusions}

\label{sec:Conclusions}

In this work we have studied the effect of magnetic dilution in a
system which hosts, in the clean limit, a topological magnon insulating
phase. We have shown that the topological phase is robust to the presence
of magnetic vacancies, surviving with a well defined Chern number
$C=1$ up to a moderate dilution $p_{1}^{*}$, which for the considered
model we estimate to be around $p_{1}^{*}\approx15\%$. For high enough
dilution $p_{2}^{*}>p_{1}^{*}$, the system enters a trivial phase
with $C=0$. For the considered model we estimate $p_{2}^{*}\approx23\%$.
Interestingly, we found $p_{1}^{*},p_{2}^{*}<p_{c}$, where $p_{c}$
is the classical percolation threshold, which for the honeycomb lattice
with nearest neighbor connections is $p_{c}\simeq30\%$ (taking into
account the second neighbor connections present in our model it is
$p_{c}\simeq64\%$).

Through an extensive characterization of the localization properties,
we have established that for $p<p_{1}^{*}$, when the system is in
the topological phase, the states that fill the clean limit gap are
localized. The system behaves effectively as a topological Chern insulator,
with Berry curvature carried by two energy regions of extended states
above and below the region of localized states. For $p>p_{2}^{*}$,
when the system is in the trivial phase with $C=0$, all states are
localized. In the crossover region, for $p_{1}^{*}<p<p_{2}^{*}$,
we have found a continuum of extended states (possibly critical) around
the middle of the spectrum, in the region previously occupied by the
localized states in the topological phase for $p<p_{1}^{*}$. Based
on the localization properties and a finite scaling analysis of the
Chern number, we conjecture that, in the thermodynamic limit, the
topological index is not well defined in this region and topological
properties are not to be expected for $p_{1}^{*}<p<p_{2}^{*}$.

Replacing magnetic atoms with non-magnetic ones in a systems hosting
a magnon Chern insulator in the clean limit puts all the three phases
within experimental reach. Although most 2D ferromagnetic honeycomb
materials, like the compounds \ce{CrX_3} (\ce{X}$\equiv$\ce{Cl},\ce{I}
and \ce{Br}~\citep{Huang2017,PhysRev.134.A433,PhysRev.137.A163,PhysRevB.3.157,PhysRevB.101.134418}),
do not exhibit a strong enough DM interaction to originate nontrivial
behavior, it is possible to mimic this interaction with the use of
an external time dependent electric field~\citep{Owerre2017} or
to increase the existing interaction via structural changes in the
compound such as strain~\citep{Koretsune2015} and the addition of
more layers~\citep{Legrand2022}, allowing the clean limit topological
properties to be engineered~\citep{Antao2023}. An interesting open
question is whether dilution in topological kagome ferromagnets leads
to qualitatively similar physics~\citep{Zhuo2021,Zhuo2022}.

\section*{Acknowledgements}
\begin{acknowledgments}
MSO and EVC acknowledge partial support from Fundação para a Ciência
e Tecnologia (FCT-Portugal) through Grant No.~UIDB/04650/2020. NMRP
acknowledges support from the Independent Research Fund Denmark (grant
No. 2032-00045B), the Danish National Research Foundation (Project
No. DNRF165) and Fundação para a Ciência e Tecnologia (FCT-Portugal)
through Grant No.~PTDC/FIS-MAC/2045/2021.
\end{acknowledgments}

\appendix

\section{Classical percolation problem}

\label{secap:Classical-percolation}

\begin{figure}
\begin{centering}
\includegraphics[width=0.95\columnwidth]{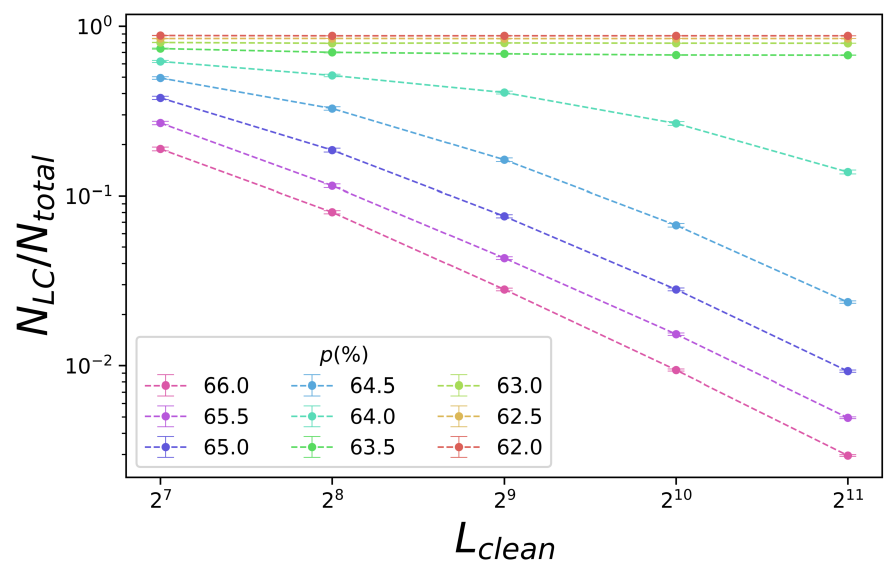}
\par\end{centering}
\caption{Fraction of sites belonging to the largest cluster as a function of
the clean system size in one spacial direction for different values
of $p$ averaged over $300$ configurations of dilution.\label{fig:Fraction-of-sites-LC}}
\end{figure}

In order to compute the classical site percolation threshold $p_{c}$
for the diluted graphs where the vertices are the honeycomb lattice
points and the first and second neighbors are the edges, we computed
the fraction of sites which belong to the largest cluster of a simulated
lattice for different clean system sizes and for different percentages
of vacancies. The results in the Fig.~\ref{fig:Fraction-of-sites-LC}
seem to indicate that for percentages below $64\%$, the fraction
of sites in the largest cluster is constant with respect to the increase
of the system size, and for percentages higher than $64\%$ there
is a clear scaling with $\propto1/N_{\text{total}}$ showing that
the largest cluster has a constant size for these percentages. We
estimate the value of $p_{c}$ to be around $64\%$ since we have
the transition from the two distinct behaviors at this point.

\section{Level spacing statistics sampling}

For each disorder configuration, we obtained the set of energies,
$e_{n}$, with exact diagonalization. With this set, we computed the
energy spacings between consecutive levels, $s_{n}=e_{n}-e_{n}$,
and used them to calculate the ratio between consecutive energy spacings
defined as $r_{n}=\min(s_{n},s_{n+1})/\max(s_{n},s_{n+1})$. This
last quantity is known to behave better in regions where the DOS is
small~\citep{Oganesyan2007,Atas2013}. After gathering the lists
of $r_{n}$ over several dilution configurations, we grouped them
in equally spaced bins over the spectrum to compute the average ratio
$\left\langle r\right\rangle $ inside each bin and obtain $\left\langle r\right\rangle $
as a function of energy and compare it with the expected result for
the GUE and Poisson distributions (Fig.~\ref{fig:Average-value-of-ratio}).
The bin size was chosen small enough so we could probe the principal
changes in localization properties along the topological transition,
but large enough to enable $\left\langle r\right\rangle $ to have
a smooth behavior.

\bibliographystyle{revtex4-1}
\bibliography{refs-topo-magnons}

\begin{thebibliography}{84}%
\makeatletter
\providecommand \@ifxundefined [1]{%
 \@ifx{#1\undefined}
}%
\providecommand \@ifnum [1]{%
 \ifnum #1\expandafter \@firstoftwo
 \else \expandafter \@secondoftwo
 \fi
}%
\providecommand \@ifx [1]{%
 \ifx #1\expandafter \@firstoftwo
 \else \expandafter \@secondoftwo
 \fi
}%
\providecommand \natexlab [1]{#1}%
\providecommand \enquote  [1]{``#1''}%
\providecommand \bibnamefont  [1]{#1}%
\providecommand \bibfnamefont [1]{#1}%
\providecommand \citenamefont [1]{#1}%
\providecommand \href@noop [0]{\@secondoftwo}%
\providecommand \href [0]{\begingroup \@sanitize@url \@href}%
\providecommand \@href[1]{\@@startlink{#1}\@@href}%
\providecommand \@@href[1]{\endgroup#1\@@endlink}%
\providecommand \@sanitize@url [0]{\catcode `\\12\catcode `\$12\catcode
  `\&12\catcode `\#12\catcode `\^12\catcode `\_12\catcode `\%12\relax}%
\providecommand \@@startlink[1]{}%
\providecommand \@@endlink[0]{}%
\providecommand \url  [0]{\begingroup\@sanitize@url \@url }%
\providecommand \@url [1]{\endgroup\@href {#1}{\urlprefix }}%
\providecommand \urlprefix  [0]{URL }%
\providecommand \Eprint [0]{\href }%
\providecommand \doibase [0]{http://dx.doi.org/}%
\providecommand \selectlanguage [0]{\@gobble}%
\providecommand \bibinfo  [0]{\@secondoftwo}%
\providecommand \bibfield  [0]{\@secondoftwo}%
\providecommand \translation [1]{[#1]}%
\providecommand \BibitemOpen [0]{}%
\providecommand \bibitemStop [0]{}%
\providecommand \bibitemNoStop [0]{.\EOS\space}%
\providecommand \EOS [0]{\spacefactor3000\relax}%
\providecommand \BibitemShut  [1]{\csname bibitem#1\endcsname}%
\let\auto@bib@innerbib\@empty
\bibitem [{\citenamefont {Rajput}\ \emph {et~al.}(2022)\citenamefont {Rajput},
  \citenamefont {Bhandari},\ and\ \citenamefont {Wadhwa}}]{Rajput2022}%
  \BibitemOpen
  \bibfield  {author} {\bibinfo {author} {\bibfnamefont {P.~J.}\ \bibnamefont
  {Rajput}}, \bibinfo {author} {\bibfnamefont {S.~U.}\ \bibnamefont
  {Bhandari}}, \ and\ \bibinfo {author} {\bibfnamefont {G.}~\bibnamefont
  {Wadhwa}},\ }\href {\doibase 10.1007/s12633-021-01643-x} {\bibfield
  {journal} {\bibinfo  {journal} {Silicon}\ } (\bibinfo {year} {2022}),\
  10.1007/s12633-021-01643-x}\BibitemShut {NoStop}%
\bibitem [{\citenamefont {Lenk}\ \emph {et~al.}(2011)\citenamefont {Lenk},
  \citenamefont {Ulrichs}, \citenamefont {Garbs},\ and\ \citenamefont
  {Münzenberg}}]{Lenk2011}%
  \BibitemOpen
  \bibfield  {author} {\bibinfo {author} {\bibfnamefont {B.}~\bibnamefont
  {Lenk}}, \bibinfo {author} {\bibfnamefont {H.}~\bibnamefont {Ulrichs}},
  \bibinfo {author} {\bibfnamefont {F.}~\bibnamefont {Garbs}}, \ and\ \bibinfo
  {author} {\bibfnamefont {M.}~\bibnamefont {Münzenberg}},\ }\href {\doibase
  10.1016/j.physrep.2011.06.003} {\bibfield  {journal} {\bibinfo  {journal}
  {Physics Reports}\ }\textbf {\bibinfo {volume} {507}},\ \bibinfo {pages}
  {107} (\bibinfo {year} {2011})}\BibitemShut {NoStop}%
\bibitem [{\citenamefont {Serga}\ \emph {et~al.}(2004)\citenamefont {Serga},
  \citenamefont {Demokritov}, \citenamefont {Hillebrands},\ and\ \citenamefont
  {Slavin}}]{Serga2004}%
  \BibitemOpen
  \bibfield  {author} {\bibinfo {author} {\bibfnamefont {A.~A.}\ \bibnamefont
  {Serga}}, \bibinfo {author} {\bibfnamefont {S.~O.}\ \bibnamefont
  {Demokritov}}, \bibinfo {author} {\bibfnamefont {B.}~\bibnamefont
  {Hillebrands}}, \ and\ \bibinfo {author} {\bibfnamefont {A.~N.}\ \bibnamefont
  {Slavin}},\ }\href {\doibase 10.1103/PhysRevLett.92.117203} {\bibfield
  {journal} {\bibinfo  {journal} {Physical Review Letters}\ }\textbf {\bibinfo
  {volume} {92}},\ \bibinfo {pages} {117203} (\bibinfo {year}
  {2004})}\BibitemShut {NoStop}%
\bibitem [{\citenamefont {Demidov}\ \emph {et~al.}(2009)\citenamefont
  {Demidov}, \citenamefont {Kostylev}, \citenamefont {Rott}, \citenamefont
  {Krzysteczko}, \citenamefont {Reiss},\ and\ \citenamefont
  {Demokritov}}]{Demidov2009}%
  \BibitemOpen
  \bibfield  {author} {\bibinfo {author} {\bibfnamefont {V.~E.}\ \bibnamefont
  {Demidov}}, \bibinfo {author} {\bibfnamefont {M.~P.}\ \bibnamefont
  {Kostylev}}, \bibinfo {author} {\bibfnamefont {K.}~\bibnamefont {Rott}},
  \bibinfo {author} {\bibfnamefont {P.}~\bibnamefont {Krzysteczko}}, \bibinfo
  {author} {\bibfnamefont {G.}~\bibnamefont {Reiss}}, \ and\ \bibinfo {author}
  {\bibfnamefont {S.~O.}\ \bibnamefont {Demokritov}},\ }\href {\doibase
  10.1063/1.3231875} {\bibfield  {journal} {\bibinfo  {journal} {Applied
  Physics Letters}\ }\textbf {\bibinfo {volume} {95}},\ \bibinfo {pages}
  {112509} (\bibinfo {year} {2009})}\BibitemShut {NoStop}%
\bibitem [{\citenamefont {Jorzick}\ \emph {et~al.}(2002)\citenamefont
  {Jorzick}, \citenamefont {Demokritov}, \citenamefont {Hillebrands},
  \citenamefont {Bailleul}, \citenamefont {Fermon}, \citenamefont {Guslienko},
  \citenamefont {Slavin}, \citenamefont {Berkov},\ and\ \citenamefont
  {Gorn}}]{Jorzick2002}%
  \BibitemOpen
  \bibfield  {author} {\bibinfo {author} {\bibfnamefont {J.}~\bibnamefont
  {Jorzick}}, \bibinfo {author} {\bibfnamefont {S.~O.}\ \bibnamefont
  {Demokritov}}, \bibinfo {author} {\bibfnamefont {B.}~\bibnamefont
  {Hillebrands}}, \bibinfo {author} {\bibfnamefont {M.}~\bibnamefont
  {Bailleul}}, \bibinfo {author} {\bibfnamefont {C.}~\bibnamefont {Fermon}},
  \bibinfo {author} {\bibfnamefont {K.~Y.}\ \bibnamefont {Guslienko}}, \bibinfo
  {author} {\bibfnamefont {A.~N.}\ \bibnamefont {Slavin}}, \bibinfo {author}
  {\bibfnamefont {D.~V.}\ \bibnamefont {Berkov}}, \ and\ \bibinfo {author}
  {\bibfnamefont {N.~L.}\ \bibnamefont {Gorn}},\ }\href {\doibase
  10.1103/PhysRevLett.88.047204} {\bibfield  {journal} {\bibinfo  {journal}
  {Physical Review Letters}\ }\textbf {\bibinfo {volume} {88}},\ \bibinfo
  {pages} {047204} (\bibinfo {year} {2002})}\BibitemShut {NoStop}%
\bibitem [{\citenamefont {Podbielski}\ \emph {et~al.}(2006)\citenamefont
  {Podbielski}, \citenamefont {Giesen},\ and\ \citenamefont
  {Grundler}}]{Podbielski2006}%
  \BibitemOpen
  \bibfield  {author} {\bibinfo {author} {\bibfnamefont {J.}~\bibnamefont
  {Podbielski}}, \bibinfo {author} {\bibfnamefont {F.}~\bibnamefont {Giesen}},
  \ and\ \bibinfo {author} {\bibfnamefont {D.}~\bibnamefont {Grundler}},\
  }\href {\doibase 10.1103/PhysRevLett.96.167207} {\bibfield  {journal}
  {\bibinfo  {journal} {Physical Review Letters}\ }\textbf {\bibinfo {volume}
  {96}},\ \bibinfo {pages} {167207} (\bibinfo {year} {2006})}\BibitemShut
  {NoStop}%
\bibitem [{\citenamefont {Barman}\ \emph {et~al.}(2021)\citenamefont {Barman},
  \citenamefont {Gubbiotti}, \citenamefont {Ladak}, \citenamefont {Adeyeye},
  \citenamefont {Krawczyk}, \citenamefont {Gräfe}, \citenamefont {Adelmann},
  \citenamefont {Cotofana}, \citenamefont {Naeemi}, \citenamefont {Vasyuchka},
  \citenamefont {Hillebrands}, \citenamefont {Nikitov}, \citenamefont {Yu},
  \citenamefont {Grundler}, \citenamefont {Sadovnikov}, \citenamefont
  {Grachev}, \citenamefont {Sheshukova}, \citenamefont {Duquesne},
  \citenamefont {Marangolo}, \citenamefont {Csaba}, \citenamefont {Porod},
  \citenamefont {Demidov}, \citenamefont {Urazhdin}, \citenamefont
  {Demokritov}, \citenamefont {Albisetti}, \citenamefont {Petti}, \citenamefont
  {Bertacco}, \citenamefont {Schultheiss}, \citenamefont {Kruglyak},
  \citenamefont {Poimanov}, \citenamefont {Sahoo}, \citenamefont {Sinha},
  \citenamefont {Yang}, \citenamefont {Münzenberg}, \citenamefont {Moriyama},
  \citenamefont {Mizukami}, \citenamefont {Landeros}, \citenamefont {Gallardo},
  \citenamefont {Carlotti}, \citenamefont {Kim}, \citenamefont {Stamps},
  \citenamefont {Camley}, \citenamefont {Rana}, \citenamefont {Otani},
  \citenamefont {Yu}, \citenamefont {Yu}, \citenamefont {Bauer}, \citenamefont
  {Back}, \citenamefont {Uhrig}, \citenamefont {Dobrovolskiy}, \citenamefont
  {Budinska}, \citenamefont {Qin}, \citenamefont {van Dijken}, \citenamefont
  {Chumak}, \citenamefont {Khitun}, \citenamefont {Nikonov}, \citenamefont
  {Young}, \citenamefont {Zingsem},\ and\ \citenamefont
  {Winklhofer}}]{Barman2021}%
  \BibitemOpen
  \bibfield  {author} {\bibinfo {author} {\bibfnamefont {A.}~\bibnamefont
  {Barman}}, \bibinfo {author} {\bibfnamefont {G.}~\bibnamefont {Gubbiotti}},
  \bibinfo {author} {\bibfnamefont {S.}~\bibnamefont {Ladak}}, \bibinfo
  {author} {\bibfnamefont {A.~O.}\ \bibnamefont {Adeyeye}}, \bibinfo {author}
  {\bibfnamefont {M.}~\bibnamefont {Krawczyk}}, \bibinfo {author}
  {\bibfnamefont {J.}~\bibnamefont {Gräfe}}, \bibinfo {author} {\bibfnamefont
  {C.}~\bibnamefont {Adelmann}}, \bibinfo {author} {\bibfnamefont
  {S.}~\bibnamefont {Cotofana}}, \bibinfo {author} {\bibfnamefont
  {A.}~\bibnamefont {Naeemi}}, \bibinfo {author} {\bibfnamefont {V.~I.}\
  \bibnamefont {Vasyuchka}}, \bibinfo {author} {\bibfnamefont {B.}~\bibnamefont
  {Hillebrands}}, \bibinfo {author} {\bibfnamefont {S.~A.}\ \bibnamefont
  {Nikitov}}, \bibinfo {author} {\bibfnamefont {H.}~\bibnamefont {Yu}},
  \bibinfo {author} {\bibfnamefont {D.}~\bibnamefont {Grundler}}, \bibinfo
  {author} {\bibfnamefont {A.~V.}\ \bibnamefont {Sadovnikov}}, \bibinfo
  {author} {\bibfnamefont {A.~A.}\ \bibnamefont {Grachev}}, \bibinfo {author}
  {\bibfnamefont {S.~E.}\ \bibnamefont {Sheshukova}}, \bibinfo {author}
  {\bibfnamefont {J.-Y.}\ \bibnamefont {Duquesne}}, \bibinfo {author}
  {\bibfnamefont {M.}~\bibnamefont {Marangolo}}, \bibinfo {author}
  {\bibfnamefont {G.}~\bibnamefont {Csaba}}, \bibinfo {author} {\bibfnamefont
  {W.}~\bibnamefont {Porod}}, \bibinfo {author} {\bibfnamefont {V.~E.}\
  \bibnamefont {Demidov}}, \bibinfo {author} {\bibfnamefont {S.}~\bibnamefont
  {Urazhdin}}, \bibinfo {author} {\bibfnamefont {S.~O.}\ \bibnamefont
  {Demokritov}}, \bibinfo {author} {\bibfnamefont {E.}~\bibnamefont
  {Albisetti}}, \bibinfo {author} {\bibfnamefont {D.}~\bibnamefont {Petti}},
  \bibinfo {author} {\bibfnamefont {R.}~\bibnamefont {Bertacco}}, \bibinfo
  {author} {\bibfnamefont {H.}~\bibnamefont {Schultheiss}}, \bibinfo {author}
  {\bibfnamefont {V.~V.}\ \bibnamefont {Kruglyak}}, \bibinfo {author}
  {\bibfnamefont {V.~D.}\ \bibnamefont {Poimanov}}, \bibinfo {author}
  {\bibfnamefont {S.}~\bibnamefont {Sahoo}}, \bibinfo {author} {\bibfnamefont
  {J.}~\bibnamefont {Sinha}}, \bibinfo {author} {\bibfnamefont
  {H.}~\bibnamefont {Yang}}, \bibinfo {author} {\bibfnamefont {M.}~\bibnamefont
  {Münzenberg}}, \bibinfo {author} {\bibfnamefont {T.}~\bibnamefont
  {Moriyama}}, \bibinfo {author} {\bibfnamefont {S.}~\bibnamefont {Mizukami}},
  \bibinfo {author} {\bibfnamefont {P.}~\bibnamefont {Landeros}}, \bibinfo
  {author} {\bibfnamefont {R.~A.}\ \bibnamefont {Gallardo}}, \bibinfo {author}
  {\bibfnamefont {G.}~\bibnamefont {Carlotti}}, \bibinfo {author}
  {\bibfnamefont {J.-V.}\ \bibnamefont {Kim}}, \bibinfo {author} {\bibfnamefont
  {R.~L.}\ \bibnamefont {Stamps}}, \bibinfo {author} {\bibfnamefont {R.~E.}\
  \bibnamefont {Camley}}, \bibinfo {author} {\bibfnamefont {B.}~\bibnamefont
  {Rana}}, \bibinfo {author} {\bibfnamefont {Y.}~\bibnamefont {Otani}},
  \bibinfo {author} {\bibfnamefont {W.}~\bibnamefont {Yu}}, \bibinfo {author}
  {\bibfnamefont {T.}~\bibnamefont {Yu}}, \bibinfo {author} {\bibfnamefont
  {G.~E.~W.}\ \bibnamefont {Bauer}}, \bibinfo {author} {\bibfnamefont
  {C.}~\bibnamefont {Back}}, \bibinfo {author} {\bibfnamefont {G.~S.}\
  \bibnamefont {Uhrig}}, \bibinfo {author} {\bibfnamefont {O.~V.}\ \bibnamefont
  {Dobrovolskiy}}, \bibinfo {author} {\bibfnamefont {B.}~\bibnamefont
  {Budinska}}, \bibinfo {author} {\bibfnamefont {H.}~\bibnamefont {Qin}},
  \bibinfo {author} {\bibfnamefont {S.}~\bibnamefont {van Dijken}}, \bibinfo
  {author} {\bibfnamefont {A.~V.}\ \bibnamefont {Chumak}}, \bibinfo {author}
  {\bibfnamefont {A.}~\bibnamefont {Khitun}}, \bibinfo {author} {\bibfnamefont
  {D.~E.}\ \bibnamefont {Nikonov}}, \bibinfo {author} {\bibfnamefont {I.~A.}\
  \bibnamefont {Young}}, \bibinfo {author} {\bibfnamefont {B.~W.}\ \bibnamefont
  {Zingsem}}, \ and\ \bibinfo {author} {\bibfnamefont {M.}~\bibnamefont
  {Winklhofer}},\ }\href {\doibase 10.1088/1361-648x/abec1a} {\bibfield
  {journal} {\bibinfo  {journal} {Journal of Physics: Condensed Matter}\
  }\textbf {\bibinfo {volume} {33}},\ \bibinfo {pages} {413001} (\bibinfo
  {year} {2021})}\BibitemShut {NoStop}%
\bibitem [{\citenamefont {Jungfleisch}\ \emph {et~al.}(2015)\citenamefont
  {Jungfleisch}, \citenamefont {Zhang}, \citenamefont {Jiang}, \citenamefont
  {Chang}, \citenamefont {Sklenar}, \citenamefont {Wu}, \citenamefont
  {Pearson}, \citenamefont {Bhattacharya}, \citenamefont {Ketterson},
  \citenamefont {Wu},\ and\ \citenamefont {Hoffmann}}]{Jungfleisch2015}%
  \BibitemOpen
  \bibfield  {author} {\bibinfo {author} {\bibfnamefont {M.~B.}\ \bibnamefont
  {Jungfleisch}}, \bibinfo {author} {\bibfnamefont {W.}~\bibnamefont {Zhang}},
  \bibinfo {author} {\bibfnamefont {W.}~\bibnamefont {Jiang}}, \bibinfo
  {author} {\bibfnamefont {H.}~\bibnamefont {Chang}}, \bibinfo {author}
  {\bibfnamefont {J.}~\bibnamefont {Sklenar}}, \bibinfo {author} {\bibfnamefont
  {S.~M.}\ \bibnamefont {Wu}}, \bibinfo {author} {\bibfnamefont {J.~E.}\
  \bibnamefont {Pearson}}, \bibinfo {author} {\bibfnamefont {A.}~\bibnamefont
  {Bhattacharya}}, \bibinfo {author} {\bibfnamefont {J.~B.}\ \bibnamefont
  {Ketterson}}, \bibinfo {author} {\bibfnamefont {M.}~\bibnamefont {Wu}}, \
  and\ \bibinfo {author} {\bibfnamefont {A.}~\bibnamefont {Hoffmann}},\ }\href
  {\doibase 10.1063/1.4916027} {\bibfield  {journal} {\bibinfo  {journal}
  {Journal of Applied Physics}\ }\textbf {\bibinfo {volume} {117}},\ \bibinfo
  {pages} {17D128} (\bibinfo {year} {2015})},\ \Eprint
  {http://arxiv.org/abs/https://doi.org/10.1063/1.4916027}
  {https://doi.org/10.1063/1.4916027} \BibitemShut {NoStop}%
\bibitem [{\citenamefont {Onose}\ \emph {et~al.}(2010)\citenamefont {Onose},
  \citenamefont {Ideue}, \citenamefont {Katsura}, \citenamefont {Shiomi},
  \citenamefont {Nagaosa},\ and\ \citenamefont {Tokura}}]{Onose2010}%
  \BibitemOpen
  \bibfield  {author} {\bibinfo {author} {\bibfnamefont {Y.}~\bibnamefont
  {Onose}}, \bibinfo {author} {\bibfnamefont {T.}~\bibnamefont {Ideue}},
  \bibinfo {author} {\bibfnamefont {H.}~\bibnamefont {Katsura}}, \bibinfo
  {author} {\bibfnamefont {Y.}~\bibnamefont {Shiomi}}, \bibinfo {author}
  {\bibfnamefont {N.}~\bibnamefont {Nagaosa}}, \ and\ \bibinfo {author}
  {\bibfnamefont {Y.}~\bibnamefont {Tokura}},\ }\href {\doibase
  10.1126/science.1188260} {\bibfield  {journal} {\bibinfo  {journal}
  {Science}\ }\textbf {\bibinfo {volume} {329}},\ \bibinfo {pages} {297}
  (\bibinfo {year} {2010})},\ \bibinfo {note} {arXiv:1008.1564
  [cond-mat]}\BibitemShut {NoStop}%
\bibitem [{\citenamefont {Hirschberger}\ \emph
  {et~al.}(2015{\natexlab{a}})\citenamefont {Hirschberger}, \citenamefont
  {Krizan}, \citenamefont {Cava},\ and\ \citenamefont {Ong}}]{PMID:25838381}%
  \BibitemOpen
  \bibfield  {author} {\bibinfo {author} {\bibfnamefont {M.}~\bibnamefont
  {Hirschberger}}, \bibinfo {author} {\bibfnamefont {J.~W.}\ \bibnamefont
  {Krizan}}, \bibinfo {author} {\bibfnamefont {R.}~\bibnamefont {Cava}}, \ and\
  \bibinfo {author} {\bibfnamefont {N.}~\bibnamefont {Ong}},\ }\href {\doibase
  10.1126/science.1257340} {\bibfield  {journal} {\bibinfo  {journal} {Science
  (New York, N.Y.)}\ }\textbf {\bibinfo {volume} {348}},\ \bibinfo {pages}
  {106—109} (\bibinfo {year} {2015}{\natexlab{a}})}\BibitemShut {NoStop}%
\bibitem [{\citenamefont {Hirschberger}\ \emph
  {et~al.}(2015{\natexlab{b}})\citenamefont {Hirschberger}, \citenamefont
  {Chisnell}, \citenamefont {Lee},\ and\ \citenamefont
  {Ong}}]{Hirschberger2015}%
  \BibitemOpen
  \bibfield  {author} {\bibinfo {author} {\bibfnamefont {M.}~\bibnamefont
  {Hirschberger}}, \bibinfo {author} {\bibfnamefont {R.}~\bibnamefont
  {Chisnell}}, \bibinfo {author} {\bibfnamefont {Y.~S.}\ \bibnamefont {Lee}}, \
  and\ \bibinfo {author} {\bibfnamefont {N.~P.}\ \bibnamefont {Ong}},\ }\href
  {\doibase 10.1103/PhysRevLett.115.106603} {\bibfield  {journal} {\bibinfo
  {journal} {Phys. Rev. Lett.}\ }\textbf {\bibinfo {volume} {115}},\ \bibinfo
  {pages} {106603} (\bibinfo {year} {2015}{\natexlab{b}})},\ \Eprint
  {http://arxiv.org/abs/1502.05688} {arXiv:1502.05688} \BibitemShut {NoStop}%
\bibitem [{\citenamefont {Chisnell}\ \emph {et~al.}(2015)\citenamefont
  {Chisnell}, \citenamefont {Helton}, \citenamefont {Freedman}, \citenamefont
  {Singh}, \citenamefont {Bewley}, \citenamefont {Nocera},\ and\ \citenamefont
  {Lee}}]{Chisnell2015}%
  \BibitemOpen
  \bibfield  {author} {\bibinfo {author} {\bibfnamefont {R.}~\bibnamefont
  {Chisnell}}, \bibinfo {author} {\bibfnamefont {J.~S.}\ \bibnamefont
  {Helton}}, \bibinfo {author} {\bibfnamefont {D.~E.}\ \bibnamefont
  {Freedman}}, \bibinfo {author} {\bibfnamefont {D.~K.}\ \bibnamefont {Singh}},
  \bibinfo {author} {\bibfnamefont {R.~I.}\ \bibnamefont {Bewley}}, \bibinfo
  {author} {\bibfnamefont {D.~G.}\ \bibnamefont {Nocera}}, \ and\ \bibinfo
  {author} {\bibfnamefont {Y.~S.}\ \bibnamefont {Lee}},\ }\href {\doibase
  10.1103/PhysRevLett.115.147201} {\bibfield  {journal} {\bibinfo  {journal}
  {Physical Review Letters}\ }\textbf {\bibinfo {volume} {115}},\ \bibinfo
  {pages} {147201} (\bibinfo {year} {2015})}\BibitemShut {NoStop}%
\bibitem [{\citenamefont {Rikken}\ and\ \citenamefont {van
  Tiggelen}(1996)}]{Rikken1996}%
  \BibitemOpen
  \bibfield  {author} {\bibinfo {author} {\bibfnamefont {G.~L. J.~A.}\
  \bibnamefont {Rikken}}\ and\ \bibinfo {author} {\bibfnamefont {B.~A.}\
  \bibnamefont {van Tiggelen}},\ }\href {\doibase 10.1038/381054a0} {\bibfield
  {journal} {\bibinfo  {journal} {Nature}\ }\textbf {\bibinfo {volume} {381}},\
  \bibinfo {pages} {54} (\bibinfo {year} {1996})}\BibitemShut {NoStop}%
\bibitem [{\citenamefont {Zhang}\ \emph {et~al.}(2010)\citenamefont {Zhang},
  \citenamefont {Ren}, \citenamefont {Wang},\ and\ \citenamefont
  {Li}}]{Zhang2010}%
  \BibitemOpen
  \bibfield  {author} {\bibinfo {author} {\bibfnamefont {L.}~\bibnamefont
  {Zhang}}, \bibinfo {author} {\bibfnamefont {J.}~\bibnamefont {Ren}}, \bibinfo
  {author} {\bibfnamefont {J.-S.}\ \bibnamefont {Wang}}, \ and\ \bibinfo
  {author} {\bibfnamefont {B.}~\bibnamefont {Li}},\ }\href {\doibase
  10.1103/PhysRevLett.105.225901} {\bibfield  {journal} {\bibinfo  {journal}
  {Physical Review Letters}\ }\textbf {\bibinfo {volume} {105}},\ \bibinfo
  {pages} {225901} (\bibinfo {year} {2010})}\BibitemShut {NoStop}%
\bibitem [{\citenamefont {Thouless}\ \emph {et~al.}(1982)\citenamefont
  {Thouless}, \citenamefont {Kohmoto}, \citenamefont {Nightingale},\ and\
  \citenamefont {den Nijs}}]{TKNN82}%
  \BibitemOpen
  \bibfield  {author} {\bibinfo {author} {\bibfnamefont {D.~J.}\ \bibnamefont
  {Thouless}}, \bibinfo {author} {\bibfnamefont {M.}~\bibnamefont {Kohmoto}},
  \bibinfo {author} {\bibfnamefont {M.~P.}\ \bibnamefont {Nightingale}}, \ and\
  \bibinfo {author} {\bibfnamefont {M.}~\bibnamefont {den Nijs}},\ }\href
  {\doibase 10.1103/PhysRevLett.49.405} {\bibfield  {journal} {\bibinfo
  {journal} {Physical Review Letters}\ }\textbf {\bibinfo {volume} {49}},\
  \bibinfo {pages} {405} (\bibinfo {year} {1982})},\ \Eprint
  {http://arxiv.org/abs/arXiv:1011.1669v3} {arXiv:arXiv:1011.1669v3}
  \BibitemShut {NoStop}%
\bibitem [{\citenamefont {Haldane}(1988)}]{PhysRevLett.61.2015}%
  \BibitemOpen
  \bibfield  {author} {\bibinfo {author} {\bibfnamefont {F.~D.~M.}\
  \bibnamefont {Haldane}},\ }\href {\doibase 10.1103/PhysRevLett.61.2015}
  {\bibfield  {journal} {\bibinfo  {journal} {Phys. Rev. Lett.}\ }\textbf
  {\bibinfo {volume} {61}},\ \bibinfo {pages} {2015} (\bibinfo {year}
  {1988})}\BibitemShut {NoStop}%
\bibitem [{\citenamefont {Kane}\ and\ \citenamefont {Mele}(2005)}]{KM05}%
  \BibitemOpen
  \bibfield  {author} {\bibinfo {author} {\bibfnamefont {C.~L.}\ \bibnamefont
  {Kane}}\ and\ \bibinfo {author} {\bibfnamefont {E.~J.}\ \bibnamefont
  {Mele}},\ }\href {\doibase 10.1103/PhysRevLett.95.226801} {\bibfield
  {journal} {\bibinfo  {journal} {Physical Review Letters}\ }\textbf {\bibinfo
  {volume} {95}},\ \bibinfo {pages} {226801} (\bibinfo {year}
  {2005})}\BibitemShut {NoStop}%
\bibitem [{\citenamefont {Fu}(2010)}]{Fu2010}%
  \BibitemOpen
  \bibfield  {author} {\bibinfo {author} {\bibfnamefont {L.}~\bibnamefont
  {Fu}},\ }\href {\doibase 10.1103/PhysRevLett.104.056402} {\bibfield
  {journal} {\bibinfo  {journal} {Physical Review Letters}\ }\textbf {\bibinfo
  {volume} {104}},\ \bibinfo {pages} {056402} (\bibinfo {year}
  {2010})}\BibitemShut {NoStop}%
\bibitem [{\citenamefont {Vijay}\ \emph {et~al.}(2015)\citenamefont {Vijay},
  \citenamefont {Hsieh},\ and\ \citenamefont {Fu}}]{Vijay2015}%
  \BibitemOpen
  \bibfield  {author} {\bibinfo {author} {\bibfnamefont {S.}~\bibnamefont
  {Vijay}}, \bibinfo {author} {\bibfnamefont {T.~H.}\ \bibnamefont {Hsieh}}, \
  and\ \bibinfo {author} {\bibfnamefont {L.}~\bibnamefont {Fu}},\ }\href
  {\doibase 10.1103/PhysRevX.5.041038} {\bibfield  {journal} {\bibinfo
  {journal} {Physical Review X}\ }\textbf {\bibinfo {volume} {5}},\ \bibinfo
  {pages} {041038} (\bibinfo {year} {2015})}\BibitemShut {NoStop}%
\bibitem [{\citenamefont {Dennis}\ \emph {et~al.}(2002)\citenamefont {Dennis},
  \citenamefont {Kitaev}, \citenamefont {Landahl},\ and\ \citenamefont
  {Preskill}}]{Dennis2002}%
  \BibitemOpen
  \bibfield  {author} {\bibinfo {author} {\bibfnamefont {E.}~\bibnamefont
  {Dennis}}, \bibinfo {author} {\bibfnamefont {A.}~\bibnamefont {Kitaev}},
  \bibinfo {author} {\bibfnamefont {A.}~\bibnamefont {Landahl}}, \ and\
  \bibinfo {author} {\bibfnamefont {J.}~\bibnamefont {Preskill}},\ }\href
  {\doibase 10.1063/1.1499754} {\bibfield  {journal} {\bibinfo  {journal}
  {Journal of Mathematical Physics}\ }\textbf {\bibinfo {volume} {43}},\
  \bibinfo {pages} {4452} (\bibinfo {year} {2002})},\ \bibinfo {note}
  {arXiv:quant-ph/0110143}\BibitemShut {NoStop}%
\bibitem [{\citenamefont {Lado}\ \emph {et~al.}(2015)\citenamefont {Lado},
  \citenamefont {Garcia-Martinez},\ and\ \citenamefont
  {Fernandez-Rossier}}]{Lado2015}%
  \BibitemOpen
  \bibfield  {author} {\bibinfo {author} {\bibfnamefont {J.~L.}\ \bibnamefont
  {Lado}}, \bibinfo {author} {\bibfnamefont {N.}~\bibnamefont
  {Garcia-Martinez}}, \ and\ \bibinfo {author} {\bibfnamefont {J.}~\bibnamefont
  {Fernandez-Rossier}},\ }\href {\doibase 10.48550/ARXIV.1502.07112} {\
  (\bibinfo {year} {2015}),\ 10.48550/ARXIV.1502.07112}\BibitemShut {NoStop}%
\bibitem [{\citenamefont {Murakami}(2011)}]{Murakami2011}%
  \BibitemOpen
  \bibfield  {author} {\bibinfo {author} {\bibfnamefont {S.}~\bibnamefont
  {Murakami}},\ }\href {\doibase 10.1088/1742-6596/302/1/012019} {\bibfield
  {journal} {\bibinfo  {journal} {Journal of Physics: Conference Series}\
  }\textbf {\bibinfo {volume} {302}},\ \bibinfo {pages} {012019} (\bibinfo
  {year} {2011})}\BibitemShut {NoStop}%
\bibitem [{\citenamefont {Moriya}(1960)}]{Moriya1960}%
  \BibitemOpen
  \bibfield  {author} {\bibinfo {author} {\bibfnamefont {T.}~\bibnamefont
  {Moriya}},\ }\href {\doibase 10.1103/PhysRev.120.91} {\bibfield  {journal}
  {\bibinfo  {journal} {Physical Review}\ }\textbf {\bibinfo {volume} {120}},\
  \bibinfo {pages} {91} (\bibinfo {year} {1960})}\BibitemShut {NoStop}%
\bibitem [{\citenamefont {McClarty}(2022)}]{McClarty2021}%
  \BibitemOpen
  \bibfield  {author} {\bibinfo {author} {\bibfnamefont {P.~A.}\ \bibnamefont
  {McClarty}},\ }\href {\doibase 10.1146/annurev-conmatphys-031620-104715}
  {\bibfield  {journal} {\bibinfo  {journal} {Annual Review of Condensed Matter
  Physics}\ }\textbf {\bibinfo {volume} {13}},\ \bibinfo {pages} {171}
  (\bibinfo {year} {2022})},\ \Eprint {http://arxiv.org/abs/2106.01430}
  {arXiv:2106.01430} \BibitemShut {NoStop}%
\bibitem [{\citenamefont {Wu}\ \emph {et~al.}(2017)\citenamefont {Wu},
  \citenamefont {Song}, \citenamefont {Zhou},\ and\ \citenamefont
  {Jiang}}]{Wu2017}%
  \BibitemOpen
  \bibfield  {author} {\bibinfo {author} {\bibfnamefont {B.}~\bibnamefont
  {Wu}}, \bibinfo {author} {\bibfnamefont {J.}~\bibnamefont {Song}}, \bibinfo
  {author} {\bibfnamefont {J.}~\bibnamefont {Zhou}}, \ and\ \bibinfo {author}
  {\bibfnamefont {H.}~\bibnamefont {Jiang}},\ }\href {\doibase
  10.48550/ARXIV.1711.10725} {\  (\bibinfo {year} {2017}),\
  10.48550/ARXIV.1711.10725}\BibitemShut {NoStop}%
\bibitem [{\citenamefont {Li}(2019)}]{Li2019}%
  \BibitemOpen
  \bibfield  {author} {\bibinfo {author} {\bibfnamefont {Y.}~\bibnamefont
  {Li}},\ }\href {\doibase 10.1038/s41567-018-0399-y} {\bibfield  {journal}
  {\bibinfo  {journal} {Nature Physics}\ }\textbf {\bibinfo {volume} {15}},\
  \bibinfo {pages} {4} (\bibinfo {year} {2019})}\BibitemShut {NoStop}%
\bibitem [{\citenamefont {Xiao}\ \emph {et~al.}(2010)\citenamefont {Xiao},
  \citenamefont {Chang},\ and\ \citenamefont {Niu}}]{Xiao2010}%
  \BibitemOpen
  \bibfield  {author} {\bibinfo {author} {\bibfnamefont {D.}~\bibnamefont
  {Xiao}}, \bibinfo {author} {\bibfnamefont {M.-C.}\ \bibnamefont {Chang}}, \
  and\ \bibinfo {author} {\bibfnamefont {Q.}~\bibnamefont {Niu}},\ }\href
  {\doibase 10.1103/revmodphys.82.1959} {\bibfield  {journal} {\bibinfo
  {journal} {Reviews of Modern Physics}\ }\textbf {\bibinfo {volume} {82}},\
  \bibinfo {pages} {1959} (\bibinfo {year} {2010})}\BibitemShut {NoStop}%
\bibitem [{\citenamefont {Kramer}\ and\ \citenamefont
  {MacKinnon}(1993)}]{Kramer1993}%
  \BibitemOpen
  \bibfield  {author} {\bibinfo {author} {\bibfnamefont {B.}~\bibnamefont
  {Kramer}}\ and\ \bibinfo {author} {\bibfnamefont {A.}~\bibnamefont
  {MacKinnon}},\ }\href {\doibase 10.1088/0034-4885/56/12/001} {\bibfield
  {journal} {\bibinfo  {journal} {Rep. Prog. Phys.}\ }\textbf {\bibinfo
  {volume} {56}},\ \bibinfo {pages} {1469} (\bibinfo {year}
  {1993})}\BibitemShut {NoStop}%
\bibitem [{\citenamefont {Groth}\ \emph {et~al.}(2009)\citenamefont {Groth},
  \citenamefont {Wimmer}, \citenamefont {Akhmerov}, \citenamefont
  {Tworzyd{\l}o},\ and\ \citenamefont {Beenakker}}]{Groth2009}%
  \BibitemOpen
  \bibfield  {author} {\bibinfo {author} {\bibfnamefont {C.~W.}\ \bibnamefont
  {Groth}}, \bibinfo {author} {\bibfnamefont {M.}~\bibnamefont {Wimmer}},
  \bibinfo {author} {\bibfnamefont {A.~R.}\ \bibnamefont {Akhmerov}}, \bibinfo
  {author} {\bibfnamefont {J.}~\bibnamefont {Tworzyd{\l}o}}, \ and\ \bibinfo
  {author} {\bibfnamefont {C.~W.~J.}\ \bibnamefont {Beenakker}},\ }\href
  {\doibase 10.1103/physrevlett.103.196805} {\bibfield  {journal} {\bibinfo
  {journal} {Physical Review Letters}\ }\textbf {\bibinfo {volume} {103}},\
  \bibinfo {pages} {196805} (\bibinfo {year} {2009})}\BibitemShut {NoStop}%
\bibitem [{\citenamefont {Li}\ \emph {et~al.}(2009)\citenamefont {Li},
  \citenamefont {Chu}, \citenamefont {Jain},\ and\ \citenamefont
  {Shen}}]{Li2009}%
  \BibitemOpen
  \bibfield  {author} {\bibinfo {author} {\bibfnamefont {J.}~\bibnamefont
  {Li}}, \bibinfo {author} {\bibfnamefont {R.-L.}\ \bibnamefont {Chu}},
  \bibinfo {author} {\bibfnamefont {J.~K.}\ \bibnamefont {Jain}}, \ and\
  \bibinfo {author} {\bibfnamefont {S.-Q.}\ \bibnamefont {Shen}},\ }\href
  {\doibase 10.1103/physrevlett.102.136806} {\bibfield  {journal} {\bibinfo
  {journal} {Physical Review Letters}\ }\textbf {\bibinfo {volume} {102}},\
  \bibinfo {pages} {136806} (\bibinfo {year} {2009})}\BibitemShut {NoStop}%
\bibitem [{\citenamefont {Gonçalves}\ \emph {et~al.}(2018)\citenamefont
  {Gonçalves}, \citenamefont {Ribeiro},\ and\ \citenamefont
  {Castro}}]{Goncalves2018}%
  \BibitemOpen
  \bibfield  {author} {\bibinfo {author} {\bibfnamefont {M.}~\bibnamefont
  {Gonçalves}}, \bibinfo {author} {\bibfnamefont {P.}~\bibnamefont {Ribeiro}},
  \ and\ \bibinfo {author} {\bibfnamefont {E.~V.}\ \bibnamefont {Castro}},\
  }\href {http://arxiv.org/abs/1807.11247} {\bibfield  {journal} {\bibinfo
  {journal} {arXiv:1807.11247 [cond-mat, physics:quant-ph]}\ } (\bibinfo {year}
  {2018})},\ \bibinfo {note} {arXiv: 1807.11247}\BibitemShut {NoStop}%
\bibitem [{\citenamefont {Meier}\ \emph {et~al.}(2018)\citenamefont {Meier},
  \citenamefont {An}, \citenamefont {Dauphin}, \citenamefont {Maffei},
  \citenamefont {Massignan}, \citenamefont {Hughes},\ and\ \citenamefont
  {Gadway}}]{Meier2018}%
  \BibitemOpen
  \bibfield  {author} {\bibinfo {author} {\bibfnamefont {E.~J.}\ \bibnamefont
  {Meier}}, \bibinfo {author} {\bibfnamefont {F.~A.}\ \bibnamefont {An}},
  \bibinfo {author} {\bibfnamefont {A.}~\bibnamefont {Dauphin}}, \bibinfo
  {author} {\bibfnamefont {M.}~\bibnamefont {Maffei}}, \bibinfo {author}
  {\bibfnamefont {P.}~\bibnamefont {Massignan}}, \bibinfo {author}
  {\bibfnamefont {T.~L.}\ \bibnamefont {Hughes}}, \ and\ \bibinfo {author}
  {\bibfnamefont {B.}~\bibnamefont {Gadway}},\ }\href {\doibase
  10.1126/science.aat3406} {\bibfield  {journal} {\bibinfo  {journal}
  {Science}\ }\textbf {\bibinfo {volume} {362}},\ \bibinfo {pages} {929}
  (\bibinfo {year} {2018})}\BibitemShut {NoStop}%
\bibitem [{\citenamefont {St{\"{u}}tzer}\ \emph {et~al.}(2018)\citenamefont
  {St{\"{u}}tzer}, \citenamefont {Plotnik}, \citenamefont {Lumer},
  \citenamefont {Titum}, \citenamefont {Lindner}, \citenamefont {Segev},
  \citenamefont {Rechtsman},\ and\ \citenamefont {Szameit}}]{Stutzer2018}%
  \BibitemOpen
  \bibfield  {author} {\bibinfo {author} {\bibfnamefont {S.}~\bibnamefont
  {St{\"{u}}tzer}}, \bibinfo {author} {\bibfnamefont {Y.}~\bibnamefont
  {Plotnik}}, \bibinfo {author} {\bibfnamefont {Y.}~\bibnamefont {Lumer}},
  \bibinfo {author} {\bibfnamefont {P.}~\bibnamefont {Titum}}, \bibinfo
  {author} {\bibfnamefont {N.~H.}\ \bibnamefont {Lindner}}, \bibinfo {author}
  {\bibfnamefont {M.}~\bibnamefont {Segev}}, \bibinfo {author} {\bibfnamefont
  {M.~C.}\ \bibnamefont {Rechtsman}}, \ and\ \bibinfo {author} {\bibfnamefont
  {A.}~\bibnamefont {Szameit}},\ }\href {\doibase 10.1038/s41586-018-0418-2}
  {\bibfield  {journal} {\bibinfo  {journal} {Nature}\ }\textbf {\bibinfo
  {volume} {560}},\ \bibinfo {pages} {461} (\bibinfo {year}
  {2018})}\BibitemShut {NoStop}%
\bibitem [{\citenamefont {Zhang}\ \emph {et~al.}(2019)\citenamefont {Zhang},
  \citenamefont {Wu}, \citenamefont {Song},\ and\ \citenamefont
  {Jiang}}]{Zhang2019}%
  \BibitemOpen
  \bibfield  {author} {\bibinfo {author} {\bibfnamefont {Z.-Q.}\ \bibnamefont
  {Zhang}}, \bibinfo {author} {\bibfnamefont {B.-L.}\ \bibnamefont {Wu}},
  \bibinfo {author} {\bibfnamefont {J.}~\bibnamefont {Song}}, \ and\ \bibinfo
  {author} {\bibfnamefont {H.}~\bibnamefont {Jiang}},\ }\href {\doibase
  10.1103/PhysRevB.100.184202} {\bibfield  {journal} {\bibinfo  {journal}
  {Physical Review B}\ }\textbf {\bibinfo {volume} {100}},\ \bibinfo {pages}
  {184202} (\bibinfo {year} {2019})},\ \Eprint
  {http://arxiv.org/abs/1906.04064} {arXiv:1906.04064} \BibitemShut {NoStop}%
\bibitem [{\citenamefont {Li}\ \emph {et~al.}(2020)\citenamefont {Li},
  \citenamefont {Fu}, \citenamefont {Hu}, \citenamefont {Li},\ and\
  \citenamefont {Shen}}]{Li2020}%
  \BibitemOpen
  \bibfield  {author} {\bibinfo {author} {\bibfnamefont {C.~A.}\ \bibnamefont
  {Li}}, \bibinfo {author} {\bibfnamefont {B.}~\bibnamefont {Fu}}, \bibinfo
  {author} {\bibfnamefont {Z.~A.}\ \bibnamefont {Hu}}, \bibinfo {author}
  {\bibfnamefont {J.}~\bibnamefont {Li}}, \ and\ \bibinfo {author}
  {\bibfnamefont {S.~Q.}\ \bibnamefont {Shen}},\ }\href {\doibase
  10.1103/PhysRevLett.125.166801} {\bibfield  {journal} {\bibinfo  {journal}
  {Physical Review Letters}\ }\textbf {\bibinfo {volume} {125}},\ \bibinfo
  {pages} {166801} (\bibinfo {year} {2020})},\ \Eprint
  {http://arxiv.org/abs/2008.00513} {2008.00513} \BibitemShut {NoStop}%
\bibitem [{\citenamefont {Yang}\ \emph
  {et~al.}(2021{\natexlab{a}})\citenamefont {Yang}, \citenamefont {Li},
  \citenamefont {Duan},\ and\ \citenamefont {Xu}}]{yang2021higher}%
  \BibitemOpen
  \bibfield  {author} {\bibinfo {author} {\bibfnamefont {Y.-B.}\ \bibnamefont
  {Yang}}, \bibinfo {author} {\bibfnamefont {K.}~\bibnamefont {Li}}, \bibinfo
  {author} {\bibfnamefont {L.-M.}\ \bibnamefont {Duan}}, \ and\ \bibinfo
  {author} {\bibfnamefont {Y.}~\bibnamefont {Xu}},\ }\href {\doibase
  10.1103/PhysRevB.103.085408} {\bibfield  {journal} {\bibinfo  {journal}
  {Physical Review B}\ }\textbf {\bibinfo {volume} {103}},\ \bibinfo {pages}
  {085408} (\bibinfo {year} {2021}{\natexlab{a}})}\BibitemShut {NoStop}%
\bibitem [{\citenamefont {Agarwala}\ \emph {et~al.}(2020)\citenamefont
  {Agarwala}, \citenamefont {Juri{\v{c}}i{\'{c}}},\ and\ \citenamefont
  {Roy}}]{Agarwala2020}%
  \BibitemOpen
  \bibfield  {author} {\bibinfo {author} {\bibfnamefont {A.}~\bibnamefont
  {Agarwala}}, \bibinfo {author} {\bibfnamefont {V.}~\bibnamefont
  {Juri{\v{c}}i{\'{c}}}}, \ and\ \bibinfo {author} {\bibfnamefont
  {B.}~\bibnamefont {Roy}},\ }\href {\doibase 10.1103/PhysRevResearch.2.012067}
  {\bibfield  {journal} {\bibinfo  {journal} {Physical Review Research}\
  }\textbf {\bibinfo {volume} {2}},\ \bibinfo {pages} {012067} (\bibinfo {year}
  {2020})}\BibitemShut {NoStop}%
\bibitem [{\citenamefont {Wang}\ \emph {et~al.}(2021)\citenamefont {Wang},
  \citenamefont {Yang}, \citenamefont {Dai},\ and\ \citenamefont
  {Xu}}]{wang2021structural}%
  \BibitemOpen
  \bibfield  {author} {\bibinfo {author} {\bibfnamefont {J.-H.}\ \bibnamefont
  {Wang}}, \bibinfo {author} {\bibfnamefont {Y.-B.}\ \bibnamefont {Yang}},
  \bibinfo {author} {\bibfnamefont {N.}~\bibnamefont {Dai}}, \ and\ \bibinfo
  {author} {\bibfnamefont {Y.}~\bibnamefont {Xu}},\ }\href {\doibase
  10.1103/PhysRevLett.126.206404} {\bibfield  {journal} {\bibinfo  {journal}
  {Physical Review Letters}\ }\textbf {\bibinfo {volume} {126}},\ \bibinfo
  {pages} {206404} (\bibinfo {year} {2021})}\BibitemShut {NoStop}%
\bibitem [{\citenamefont {Peng}\ \emph {et~al.}(2022)\citenamefont {Peng},
  \citenamefont {Hua}, \citenamefont {Chen}, \citenamefont {Liu}, \citenamefont
  {Huang},\ and\ \citenamefont {Zhou}}]{Peng2022}%
  \BibitemOpen
  \bibfield  {author} {\bibinfo {author} {\bibfnamefont {T.}~\bibnamefont
  {Peng}}, \bibinfo {author} {\bibfnamefont {C.-B.}\ \bibnamefont {Hua}},
  \bibinfo {author} {\bibfnamefont {R.}~\bibnamefont {Chen}}, \bibinfo {author}
  {\bibfnamefont {Z.-R.}\ \bibnamefont {Liu}}, \bibinfo {author} {\bibfnamefont
  {H.-M.}\ \bibnamefont {Huang}}, \ and\ \bibinfo {author} {\bibfnamefont
  {B.}~\bibnamefont {Zhou}},\ }\href {\doibase 10.1103/PhysRevB.106.125310}
  {\bibfield  {journal} {\bibinfo  {journal} {Physical Review B}\ }\textbf
  {\bibinfo {volume} {106}},\ \bibinfo {pages} {125310} (\bibinfo {year}
  {2022})}\BibitemShut {NoStop}%
\bibitem [{\citenamefont {L{\'{o}}io}\ \emph {et~al.}(2023)\citenamefont
  {L{\'{o}}io}, \citenamefont {Gon{\c{c}}alves}, \citenamefont {Ribeiro},\ and\
  \citenamefont {Castro}}]{Loio2023}%
  \BibitemOpen
  \bibfield  {author} {\bibinfo {author} {\bibfnamefont {H.}~\bibnamefont
  {L{\'{o}}io}}, \bibinfo {author} {\bibfnamefont {M.}~\bibnamefont
  {Gon{\c{c}}alves}}, \bibinfo {author} {\bibfnamefont {P.}~\bibnamefont
  {Ribeiro}}, \ and\ \bibinfo {author} {\bibfnamefont {E.~V.}\ \bibnamefont
  {Castro}},\ }\href {http://arxiv.org/abs/2305.19209} {\  (\bibinfo {year}
  {2023})},\ \Eprint {http://arxiv.org/abs/2305.19209} {arXiv:2305.19209}
  \BibitemShut {NoStop}%
\bibitem [{\citenamefont {Zhang}\ \emph {et~al.}(2020)\citenamefont {Zhang},
  \citenamefont {Zou}, \citenamefont {Pei}, \citenamefont {He}, \citenamefont
  {Bao}, \citenamefont {Sun},\ and\ \citenamefont {Zhang}}]{Zhang2020}%
  \BibitemOpen
  \bibfield  {author} {\bibinfo {author} {\bibfnamefont {W.}~\bibnamefont
  {Zhang}}, \bibinfo {author} {\bibfnamefont {D.}~\bibnamefont {Zou}}, \bibinfo
  {author} {\bibfnamefont {Q.}~\bibnamefont {Pei}}, \bibinfo {author}
  {\bibfnamefont {W.}~\bibnamefont {He}}, \bibinfo {author} {\bibfnamefont
  {J.}~\bibnamefont {Bao}}, \bibinfo {author} {\bibfnamefont {H.}~\bibnamefont
  {Sun}}, \ and\ \bibinfo {author} {\bibfnamefont {X.}~\bibnamefont {Zhang}},\
  }\href {\doibase 10.1103/PhysRevLett.126.146802} {\bibfield  {journal}
  {\bibinfo  {journal} {Physical Review Letters}\ }\textbf {\bibinfo {volume}
  {126}},\ \bibinfo {pages} {146802} (\bibinfo {year} {2020})},\ \Eprint
  {http://arxiv.org/abs/2008.00423} {arXiv:2008.00423} \BibitemShut {NoStop}%
\bibitem [{\citenamefont {Edwards}\ and\ \citenamefont
  {Jones}(1971)}]{Edwards1971}%
  \BibitemOpen
  \bibfield  {author} {\bibinfo {author} {\bibfnamefont {S.~F.}\ \bibnamefont
  {Edwards}}\ and\ \bibinfo {author} {\bibfnamefont {R.~C.}\ \bibnamefont
  {Jones}},\ }\href {\doibase 10.1088/0022-3719/4/14/026} {\bibfield  {journal}
  {\bibinfo  {journal} {Journal of Physics C: Solid State Physics}\ }\textbf
  {\bibinfo {volume} {4}},\ \bibinfo {pages} {2109} (\bibinfo {year}
  {1971})}\BibitemShut {NoStop}%
\bibitem [{\citenamefont {Wan}\ \emph {et~al.}(1991)\citenamefont {Wan},
  \citenamefont {Harris},\ and\ \citenamefont {Adler}}]{Wan1991}%
  \BibitemOpen
  \bibfield  {author} {\bibinfo {author} {\bibfnamefont {C.~C.}\ \bibnamefont
  {Wan}}, \bibinfo {author} {\bibfnamefont {A.~B.}\ \bibnamefont {Harris}}, \
  and\ \bibinfo {author} {\bibfnamefont {J.}~\bibnamefont {Adler}},\ }\href
  {\doibase 10.1063/1.348123} {\bibfield  {journal} {\bibinfo  {journal}
  {Journal of Applied Physics}\ }\textbf {\bibinfo {volume} {69}},\ \bibinfo
  {pages} {5191} (\bibinfo {year} {1991})}\BibitemShut {NoStop}%
\bibitem [{\citenamefont {Sandvik}(2001)}]{Sandvik2001}%
  \BibitemOpen
  \bibfield  {author} {\bibinfo {author} {\bibfnamefont {A.~W.}\ \bibnamefont
  {Sandvik}},\ }\href {\doibase 10.1103/PhysRevLett.86.3209} {\bibfield
  {journal} {\bibinfo  {journal} {Physical Review Letters}\ }\textbf {\bibinfo
  {volume} {86}},\ \bibinfo {pages} {3209} (\bibinfo {year}
  {2001})}\BibitemShut {NoStop}%
\bibitem [{\citenamefont {Sandvik}(2002)}]{Sandvik2002}%
  \BibitemOpen
  \bibfield  {author} {\bibinfo {author} {\bibfnamefont {A.~W.}\ \bibnamefont
  {Sandvik}},\ }\href {\doibase 10.1103/PhysRevB.66.024418} {\bibfield
  {journal} {\bibinfo  {journal} {Physical Review B}\ }\textbf {\bibinfo
  {volume} {66}},\ \bibinfo {pages} {024418} (\bibinfo {year}
  {2002})}\BibitemShut {NoStop}%
\bibitem [{\citenamefont {Vojta}\ and\ \citenamefont
  {Schmalian}(2005)}]{Vojta2005}%
  \BibitemOpen
  \bibfield  {author} {\bibinfo {author} {\bibfnamefont {T.}~\bibnamefont
  {Vojta}}\ and\ \bibinfo {author} {\bibfnamefont {J.}~\bibnamefont
  {Schmalian}},\ }\href {\doibase 10.1103/PhysRevLett.95.237206} {\bibfield
  {journal} {\bibinfo  {journal} {Physical Review Letters}\ }\textbf {\bibinfo
  {volume} {95}},\ \bibinfo {pages} {237206} (\bibinfo {year} {2005})},\
  \Eprint {http://arxiv.org/abs/0508211} {arXiv:0508211} \BibitemShut {NoStop}%
\bibitem [{\citenamefont {Abrahams}\ \emph {et~al.}(1979)\citenamefont
  {Abrahams}, \citenamefont {Anderson}, \citenamefont {Licciardello},\ and\
  \citenamefont {Ramakrishnan}}]{Abrahams1979}%
  \BibitemOpen
  \bibfield  {author} {\bibinfo {author} {\bibfnamefont {E.}~\bibnamefont
  {Abrahams}}, \bibinfo {author} {\bibfnamefont {P.~W.}\ \bibnamefont
  {Anderson}}, \bibinfo {author} {\bibfnamefont {D.~C.}\ \bibnamefont
  {Licciardello}}, \ and\ \bibinfo {author} {\bibfnamefont {T.~V.}\
  \bibnamefont {Ramakrishnan}},\ }\href {\doibase 10.1103/PhysRevLett.42.673}
  {\bibfield  {journal} {\bibinfo  {journal} {Physical Review Letters}\
  }\textbf {\bibinfo {volume} {42}},\ \bibinfo {pages} {673} (\bibinfo {year}
  {1979})}\BibitemShut {NoStop}%
\bibitem [{\citenamefont {Anderson}(1958)}]{Anderson1958}%
  \BibitemOpen
  \bibfield  {author} {\bibinfo {author} {\bibfnamefont {P.~W.}\ \bibnamefont
  {Anderson}},\ }\href {\doibase 10.1103/physrev.109.1492} {\bibfield
  {journal} {\bibinfo  {journal} {Physical Review}\ }\textbf {\bibinfo {volume}
  {109}},\ \bibinfo {pages} {1492} (\bibinfo {year} {1958})}\BibitemShut
  {NoStop}%
\bibitem [{\citenamefont {Islam}\ and\ \citenamefont
  {Nakanishi}(2008)}]{Islam2008}%
  \BibitemOpen
  \bibfield  {author} {\bibinfo {author} {\bibfnamefont {M.}~\bibnamefont
  {Islam}}\ and\ \bibinfo {author} {\bibfnamefont {H.}~\bibnamefont
  {Nakanishi}},\ }\href {\doibase 10.1103/PhysRevE.77.061109} {\bibfield
  {journal} {\bibinfo  {journal} {Physical review. E, Statistical, nonlinear,
  and soft matter physics}\ }\textbf {\bibinfo {volume} {77}},\ \bibinfo
  {pages} {061109} (\bibinfo {year} {2008})}\BibitemShut {NoStop}%
\bibitem [{\citenamefont {Dillon}\ and\ \citenamefont
  {Nakanishi}(2014)}]{Dillon_2014}%
  \BibitemOpen
  \bibfield  {author} {\bibinfo {author} {\bibfnamefont {B.~S.}\ \bibnamefont
  {Dillon}}\ and\ \bibinfo {author} {\bibfnamefont {H.}~\bibnamefont
  {Nakanishi}},\ }\href {\doibase 10.1140/epjb/e2014-50397-4} {\bibfield
  {journal} {\bibinfo  {journal} {The European Physical Journal B}\ }\textbf
  {\bibinfo {volume} {87}} (\bibinfo {year} {2014}),\
  10.1140/epjb/e2014-50397-4}\BibitemShut {NoStop}%
\bibitem [{\citenamefont {Holstein}\ and\ \citenamefont
  {Primakoff}(1940)}]{Holstein1940}%
  \BibitemOpen
  \bibfield  {author} {\bibinfo {author} {\bibfnamefont {T.}~\bibnamefont
  {Holstein}}\ and\ \bibinfo {author} {\bibfnamefont {H.}~\bibnamefont
  {Primakoff}},\ }\href {\doibase 10.1103/PhysRev.58.1098} {\bibfield
  {journal} {\bibinfo  {journal} {Physical Review}\ }\textbf {\bibinfo {volume}
  {58}},\ \bibinfo {pages} {1098} (\bibinfo {year} {1940})}\BibitemShut
  {NoStop}%
\bibitem [{\citenamefont {Shindou}\ \emph {et~al.}(2013)\citenamefont
  {Shindou}, \citenamefont {Matsumoto}, \citenamefont {Murakami},\ and\
  \citenamefont {Ohe}}]{Shindou2013}%
  \BibitemOpen
  \bibfield  {author} {\bibinfo {author} {\bibfnamefont {R.}~\bibnamefont
  {Shindou}}, \bibinfo {author} {\bibfnamefont {R.}~\bibnamefont {Matsumoto}},
  \bibinfo {author} {\bibfnamefont {S.}~\bibnamefont {Murakami}}, \ and\
  \bibinfo {author} {\bibfnamefont {J.-i.}\ \bibnamefont {Ohe}},\ }\href
  {\doibase 10.1103/PhysRevB.87.174427} {\bibfield  {journal} {\bibinfo
  {journal} {Physical Review B}\ }\textbf {\bibinfo {volume} {87}},\ \bibinfo
  {pages} {174427} (\bibinfo {year} {2013})}\BibitemShut {NoStop}%
\bibitem [{\citenamefont {Zhang}\ \emph
  {et~al.}(2013{\natexlab{a}})\citenamefont {Zhang}, \citenamefont {Ren},
  \citenamefont {Wang},\ and\ \citenamefont {Li}}]{Zhang2013b}%
  \BibitemOpen
  \bibfield  {author} {\bibinfo {author} {\bibfnamefont {L.}~\bibnamefont
  {Zhang}}, \bibinfo {author} {\bibfnamefont {J.}~\bibnamefont {Ren}}, \bibinfo
  {author} {\bibfnamefont {J.-S.}\ \bibnamefont {Wang}}, \ and\ \bibinfo
  {author} {\bibfnamefont {B.}~\bibnamefont {Li}},\ }\href {\doibase
  10.1103/PhysRevB.87.144101} {\bibfield  {journal} {\bibinfo  {journal}
  {Physical Review B}\ }\textbf {\bibinfo {volume} {87}},\ \bibinfo {pages}
  {144101} (\bibinfo {year} {2013}{\natexlab{a}})}\BibitemShut {NoStop}%
\bibitem [{\citenamefont {Mook}\ \emph {et~al.}(2014)\citenamefont {Mook},
  \citenamefont {Henk},\ and\ \citenamefont {Mertig}}]{Mook2014}%
  \BibitemOpen
  \bibfield  {author} {\bibinfo {author} {\bibfnamefont {A.}~\bibnamefont
  {Mook}}, \bibinfo {author} {\bibfnamefont {J.}~\bibnamefont {Henk}}, \ and\
  \bibinfo {author} {\bibfnamefont {I.}~\bibnamefont {Mertig}},\ }\href
  {\doibase 10.1103/PhysRevB.90.024412} {\bibfield  {journal} {\bibinfo
  {journal} {Physical Review B}\ }\textbf {\bibinfo {volume} {90}},\ \bibinfo
  {pages} {024412} (\bibinfo {year} {2014})}\BibitemShut {NoStop}%
\bibitem [{\citenamefont {Suding}\ and\ \citenamefont
  {Ziff}(1999)}]{Suding1999}%
  \BibitemOpen
  \bibfield  {author} {\bibinfo {author} {\bibfnamefont {P.~N.}\ \bibnamefont
  {Suding}}\ and\ \bibinfo {author} {\bibfnamefont {R.~M.}\ \bibnamefont
  {Ziff}},\ }\href {\doibase 10.1103/PhysRevE.60.275} {\bibfield  {journal}
  {\bibinfo  {journal} {Physical Review E}\ }\textbf {\bibinfo {volume} {60}},\
  \bibinfo {pages} {275} (\bibinfo {year} {1999})}\BibitemShut {NoStop}%
\bibitem [{\citenamefont {Ziff}(1992)}]{Ziff1992}%
  \BibitemOpen
  \bibfield  {author} {\bibinfo {author} {\bibfnamefont {R.~M.}\ \bibnamefont
  {Ziff}},\ }\href {\doibase 10.1103/PhysRevLett.69.2670} {\bibfield  {journal}
  {\bibinfo  {journal} {Physical Review Letters}\ }\textbf {\bibinfo {volume}
  {69}},\ \bibinfo {pages} {2670} (\bibinfo {year} {1992})}\BibitemShut
  {NoStop}%
\bibitem [{\citenamefont {Fukui}\ \emph {et~al.}(2005)\citenamefont {Fukui},
  \citenamefont {Hatsugai},\ and\ \citenamefont {Suzuki}}]{Fukui2005}%
  \BibitemOpen
  \bibfield  {author} {\bibinfo {author} {\bibfnamefont {T.}~\bibnamefont
  {Fukui}}, \bibinfo {author} {\bibfnamefont {Y.}~\bibnamefont {Hatsugai}}, \
  and\ \bibinfo {author} {\bibfnamefont {H.}~\bibnamefont {Suzuki}},\ }\href
  {\doibase 10.1143/JPSJ.74.1674} {\bibfield  {journal} {\bibinfo  {journal}
  {Journal of the Physical Society of Japan}\ }\textbf {\bibinfo {volume}
  {74}},\ \bibinfo {pages} {1674} (\bibinfo {year} {2005})}\BibitemShut
  {NoStop}%
\bibitem [{\citenamefont {Zhang}\ \emph
  {et~al.}(2013{\natexlab{b}})\citenamefont {Zhang}, \citenamefont {Yang},
  \citenamefont {Ju}, \citenamefont {Sheng}, \citenamefont {Sheng},
  \citenamefont {Shen},\ and\ \citenamefont {Xing}}]{Zhang2013}%
  \BibitemOpen
  \bibfield  {author} {\bibinfo {author} {\bibfnamefont {Y.~F.}\ \bibnamefont
  {Zhang}}, \bibinfo {author} {\bibfnamefont {Y.~Y.}\ \bibnamefont {Yang}},
  \bibinfo {author} {\bibfnamefont {Y.}~\bibnamefont {Ju}}, \bibinfo {author}
  {\bibfnamefont {L.}~\bibnamefont {Sheng}}, \bibinfo {author} {\bibfnamefont
  {D.~N.}\ \bibnamefont {Sheng}}, \bibinfo {author} {\bibfnamefont
  {R.}~\bibnamefont {Shen}}, \ and\ \bibinfo {author} {\bibfnamefont {D.~Y.}\
  \bibnamefont {Xing}},\ }\href {\doibase 10.1088/1674-1056/22/11/117312}
  {\bibfield  {journal} {\bibinfo  {journal} {Chinese Physics B}\ }\textbf
  {\bibinfo {volume} {22}},\ \bibinfo {pages} {117312} (\bibinfo {year}
  {2013}{\natexlab{b}})},\ \bibinfo {note} {arXiv: 1212.6295}\BibitemShut
  {NoStop}%
\bibitem [{\citenamefont {Weiße}\ \emph {et~al.}(2006)\citenamefont {Weiße},
  \citenamefont {Wellein}, \citenamefont {Alvermann},\ and\ \citenamefont
  {Fehske}}]{Weisse2006}%
  \BibitemOpen
  \bibfield  {author} {\bibinfo {author} {\bibfnamefont {A.}~\bibnamefont
  {Weiße}}, \bibinfo {author} {\bibfnamefont {G.}~\bibnamefont {Wellein}},
  \bibinfo {author} {\bibfnamefont {A.}~\bibnamefont {Alvermann}}, \ and\
  \bibinfo {author} {\bibfnamefont {H.}~\bibnamefont {Fehske}},\ }\href
  {\doibase 10.1103/RevModPhys.78.275} {\bibfield  {journal} {\bibinfo
  {journal} {Reviews of Modern Physics}\ }\textbf {\bibinfo {volume} {78}},\
  \bibinfo {pages} {275} (\bibinfo {year} {2006})}\BibitemShut {NoStop}%
\bibitem [{\citenamefont {Evers}\ and\ \citenamefont
  {Mirlin}(2008)}]{Evers2008}%
  \BibitemOpen
  \bibfield  {author} {\bibinfo {author} {\bibfnamefont {F.}~\bibnamefont
  {Evers}}\ and\ \bibinfo {author} {\bibfnamefont {A.~D.}\ \bibnamefont
  {Mirlin}},\ }\href {\doibase 10.1103/revmodphys.80.1355} {\bibfield
  {journal} {\bibinfo  {journal} {Reviews of Modern Physics}\ }\textbf
  {\bibinfo {volume} {80}},\ \bibinfo {pages} {1355} (\bibinfo {year}
  {2008})}\BibitemShut {NoStop}%
\bibitem [{\citenamefont {Oganesyan}\ and\ \citenamefont
  {Huse}(2007)}]{Oganesyan2007}%
  \BibitemOpen
  \bibfield  {author} {\bibinfo {author} {\bibfnamefont {V.}~\bibnamefont
  {Oganesyan}}\ and\ \bibinfo {author} {\bibfnamefont {D.~A.}\ \bibnamefont
  {Huse}},\ }\href {\doibase 10.1103/physrevb.75.155111} {\bibfield  {journal}
  {\bibinfo  {journal} {Physical Review B}\ }\textbf {\bibinfo {volume} {75}},\
  \bibinfo {pages} {155111} (\bibinfo {year} {2007})}\BibitemShut {NoStop}%
\bibitem [{\citenamefont {Atas}\ \emph {et~al.}(2013)\citenamefont {Atas},
  \citenamefont {Bogomolny}, \citenamefont {Giraud},\ and\ \citenamefont
  {Roux}}]{Atas2013}%
  \BibitemOpen
  \bibfield  {author} {\bibinfo {author} {\bibfnamefont {Y.~Y.}\ \bibnamefont
  {Atas}}, \bibinfo {author} {\bibfnamefont {E.}~\bibnamefont {Bogomolny}},
  \bibinfo {author} {\bibfnamefont {O.}~\bibnamefont {Giraud}}, \ and\ \bibinfo
  {author} {\bibfnamefont {G.}~\bibnamefont {Roux}},\ }\href {\doibase
  10.1103/physrevlett.110.084101} {\bibfield  {journal} {\bibinfo  {journal}
  {Physical Review Letters}\ }\textbf {\bibinfo {volume} {110}},\ \bibinfo
  {pages} {084101} (\bibinfo {year} {2013})}\BibitemShut {NoStop}%
\bibitem [{\citenamefont {Wang}\ \emph {et~al.}(2014)\citenamefont {Wang},
  \citenamefont {Avishai}, \citenamefont {Meir},\ and\ \citenamefont
  {Wang}}]{wang2014anti}%
  \BibitemOpen
  \bibfield  {author} {\bibinfo {author} {\bibfnamefont {C.}~\bibnamefont
  {Wang}}, \bibinfo {author} {\bibfnamefont {Y.}~\bibnamefont {Avishai}},
  \bibinfo {author} {\bibfnamefont {Y.}~\bibnamefont {Meir}}, \ and\ \bibinfo
  {author} {\bibfnamefont {X.~R.}\ \bibnamefont {Wang}},\ }\href@noop {}
  {\bibfield  {journal} {\bibinfo  {journal} {Physical Review B}\ }\textbf
  {\bibinfo {volume} {89}},\ \bibinfo {pages} {45314} (\bibinfo {year}
  {2014})}\BibitemShut {NoStop}%
\bibitem [{\citenamefont {Onoda}\ and\ \citenamefont
  {Nagaosa}(2003)}]{Onoda2003}%
  \BibitemOpen
  \bibfield  {author} {\bibinfo {author} {\bibfnamefont {M.}~\bibnamefont
  {Onoda}}\ and\ \bibinfo {author} {\bibfnamefont {N.}~\bibnamefont
  {Nagaosa}},\ }\href {\doibase 10.1103/PhysRevLett.90.206601} {\bibfield
  {journal} {\bibinfo  {journal} {Physical Review Letters}\ }\textbf {\bibinfo
  {volume} {90}},\ \bibinfo {pages} {206601} (\bibinfo {year}
  {2003})}\BibitemShut {NoStop}%
\bibitem [{\citenamefont {Onoda}\ \emph {et~al.}(2007)\citenamefont {Onoda},
  \citenamefont {Avishai},\ and\ \citenamefont {Nagaosa}}]{Onoda2007}%
  \BibitemOpen
  \bibfield  {author} {\bibinfo {author} {\bibfnamefont {M.}~\bibnamefont
  {Onoda}}, \bibinfo {author} {\bibfnamefont {Y.}~\bibnamefont {Avishai}}, \
  and\ \bibinfo {author} {\bibfnamefont {N.}~\bibnamefont {Nagaosa}},\ }\href
  {\doibase 10.1103/PhysRevLett.98.076802} {\bibfield  {journal} {\bibinfo
  {journal} {Phys. Rev. Lett.}\ }\textbf {\bibinfo {volume} {98}} (\bibinfo
  {year} {2007}),\ 10.1103/PhysRevLett.98.076802},\ \Eprint
  {http://arxiv.org/abs/0605510} {0605510} \BibitemShut {NoStop}%
\bibitem [{\citenamefont {Castro}\ \emph {et~al.}(2016)\citenamefont {Castro},
  \citenamefont {de~Gail}, \citenamefont {L{\'{o}}pez-Sancho},\ and\
  \citenamefont {Vozmediano}}]{Castro2016}%
  \BibitemOpen
  \bibfield  {author} {\bibinfo {author} {\bibfnamefont {E.~V.}\ \bibnamefont
  {Castro}}, \bibinfo {author} {\bibfnamefont {R.}~\bibnamefont {de~Gail}},
  \bibinfo {author} {\bibfnamefont {M.~P.}\ \bibnamefont {L{\'{o}}pez-Sancho}},
  \ and\ \bibinfo {author} {\bibfnamefont {M.~A.~H.}\ \bibnamefont
  {Vozmediano}},\ }\href {\doibase 10.1103/PhysRevB.93.245414} {\bibfield
  {journal} {\bibinfo  {journal} {Physical Review B}\ }\textbf {\bibinfo
  {volume} {93}},\ \bibinfo {pages} {245414} (\bibinfo {year}
  {2016})}\BibitemShut {NoStop}%
\bibitem [{\citenamefont {Laughlin}(1984)}]{Laughlin1984}%
  \BibitemOpen
  \bibfield  {author} {\bibinfo {author} {\bibfnamefont {R.~B.}\ \bibnamefont
  {Laughlin}},\ }\href {\doibase 10.1103/PhysRevLett.52.2304} {\bibfield
  {journal} {\bibinfo  {journal} {Physical Review Letters}\ }\textbf {\bibinfo
  {volume} {52}},\ \bibinfo {pages} {2304} (\bibinfo {year}
  {1984})}\BibitemShut {NoStop}%
\bibitem [{\citenamefont {Prodan}(2011)}]{Prodan2011}%
  \BibitemOpen
  \bibfield  {author} {\bibinfo {author} {\bibfnamefont {E.}~\bibnamefont
  {Prodan}},\ }\href {\doibase 10.1088/1751-8113/44/11/113001} {\bibfield
  {journal} {\bibinfo  {journal} {Journal of Physics A: Mathematical and
  Theoretical}\ }\textbf {\bibinfo {volume} {44}},\ \bibinfo {pages} {113001}
  (\bibinfo {year} {2011})},\ \Eprint {http://arxiv.org/abs/1010.0595}
  {arXiv:1010.0595} \BibitemShut {NoStop}%
\bibitem [{\citenamefont {Wang}\ \emph {et~al.}(2015)\citenamefont {Wang},
  \citenamefont {Su}, \citenamefont {Avishai}, \citenamefont {Meir},\ and\
  \citenamefont {Wang}}]{avishai2015criticalMetal}%
  \BibitemOpen
  \bibfield  {author} {\bibinfo {author} {\bibfnamefont {C.}~\bibnamefont
  {Wang}}, \bibinfo {author} {\bibfnamefont {Y.}~\bibnamefont {Su}}, \bibinfo
  {author} {\bibfnamefont {Y.}~\bibnamefont {Avishai}}, \bibinfo {author}
  {\bibfnamefont {Y.}~\bibnamefont {Meir}}, \ and\ \bibinfo {author}
  {\bibfnamefont {X.~R.}\ \bibnamefont {Wang}},\ }\href {\doibase
  10.1103/PhysRevLett.114.096803} {\bibfield  {journal} {\bibinfo  {journal}
  {Physical Review Letters}\ }\textbf {\bibinfo {volume} {114}},\ \bibinfo
  {pages} {096803} (\bibinfo {year} {2015})},\ \Eprint
  {http://arxiv.org/abs/1411.4838} {arXiv:1411.4838} \BibitemShut {NoStop}%
\bibitem [{\citenamefont {Qiao}\ \emph {et~al.}(2016)\citenamefont {Qiao},
  \citenamefont {Han}, \citenamefont {Zhang}, \citenamefont {Wang},
  \citenamefont {Deng}, \citenamefont {Jiang}, \citenamefont {Yang},
  \citenamefont {Wang},\ and\ \citenamefont {Niu}}]{Qiao2016}%
  \BibitemOpen
  \bibfield  {author} {\bibinfo {author} {\bibfnamefont {Z.}~\bibnamefont
  {Qiao}}, \bibinfo {author} {\bibfnamefont {Y.}~\bibnamefont {Han}}, \bibinfo
  {author} {\bibfnamefont {L.}~\bibnamefont {Zhang}}, \bibinfo {author}
  {\bibfnamefont {K.}~\bibnamefont {Wang}}, \bibinfo {author} {\bibfnamefont
  {X.}~\bibnamefont {Deng}}, \bibinfo {author} {\bibfnamefont {H.}~\bibnamefont
  {Jiang}}, \bibinfo {author} {\bibfnamefont {S.~A.}\ \bibnamefont {Yang}},
  \bibinfo {author} {\bibfnamefont {J.}~\bibnamefont {Wang}}, \ and\ \bibinfo
  {author} {\bibfnamefont {Q.}~\bibnamefont {Niu}},\ }\href {\doibase
  10.1103/physrevlett.117.056802} {\bibfield  {journal} {\bibinfo  {journal}
  {Phys. Rev. Lett.}\ }\textbf {\bibinfo {volume} {117}},\ \bibinfo {pages}
  {056802} (\bibinfo {year} {2016})}\BibitemShut {NoStop}%
\bibitem [{\citenamefont {Su}\ \emph {et~al.}(2016)\citenamefont {Su},
  \citenamefont {Wang}, \citenamefont {Avishai}, \citenamefont {Meir},\ and\
  \citenamefont {Wang}}]{avishai2016Metal}%
  \BibitemOpen
  \bibfield  {author} {\bibinfo {author} {\bibfnamefont {Y.}~\bibnamefont
  {Su}}, \bibinfo {author} {\bibfnamefont {C.}~\bibnamefont {Wang}}, \bibinfo
  {author} {\bibfnamefont {Y.}~\bibnamefont {Avishai}}, \bibinfo {author}
  {\bibfnamefont {Y.}~\bibnamefont {Meir}}, \ and\ \bibinfo {author}
  {\bibfnamefont {X.~R.}\ \bibnamefont {Wang}},\ }\href {\doibase
  10.1038/srep33304} {\bibfield  {journal} {\bibinfo  {journal} {Scientific
  Reports}\ }\textbf {\bibinfo {volume} {6}},\ \bibinfo {pages} {33304}
  (\bibinfo {year} {2016})}\BibitemShut {NoStop}%
\bibitem [{\citenamefont {Yang}\ \emph
  {et~al.}(2021{\natexlab{b}})\citenamefont {Yang}, \citenamefont {Zeng},
  \citenamefont {Han},\ and\ \citenamefont {Qiao}}]{Yang2021}%
  \BibitemOpen
  \bibfield  {author} {\bibinfo {author} {\bibfnamefont {H.}~\bibnamefont
  {Yang}}, \bibinfo {author} {\bibfnamefont {J.}~\bibnamefont {Zeng}}, \bibinfo
  {author} {\bibfnamefont {Y.}~\bibnamefont {Han}}, \ and\ \bibinfo {author}
  {\bibfnamefont {Z.}~\bibnamefont {Qiao}},\ }\href {\doibase
  10.1103/physrevb.104.115414} {\bibfield  {journal} {\bibinfo  {journal}
  {Physical Review B}\ }\textbf {\bibinfo {volume} {104}},\ \bibinfo {pages}
  {115414} (\bibinfo {year} {2021}{\natexlab{b}})}\BibitemShut {NoStop}%
\bibitem [{\citenamefont {{Castro Neto}}\ \emph {et~al.}(2009)\citenamefont
  {{Castro Neto}}, \citenamefont {Guinea}, \citenamefont {Peres}, \citenamefont
  {Novoselov},\ and\ \citenamefont {Geim}}]{NGPrmp}%
  \BibitemOpen
  \bibfield  {author} {\bibinfo {author} {\bibfnamefont {A.~H.}\ \bibnamefont
  {{Castro Neto}}}, \bibinfo {author} {\bibfnamefont {F.}~\bibnamefont
  {Guinea}}, \bibinfo {author} {\bibfnamefont {N.~M.~R.}\ \bibnamefont
  {Peres}}, \bibinfo {author} {\bibfnamefont {K.~S.}\ \bibnamefont
  {Novoselov}}, \ and\ \bibinfo {author} {\bibfnamefont {A.~K.}\ \bibnamefont
  {Geim}},\ }\href {\doibase 10.1103/RevModPhys.81.109} {\bibfield  {journal}
  {\bibinfo  {journal} {Reviews of Modern Physics}\ }\textbf {\bibinfo {volume}
  {81}},\ \bibinfo {pages} {109} (\bibinfo {year} {2009})},\ \Eprint
  {http://arxiv.org/abs/0709.1163} {arXiv:0709.1163} \BibitemShut {NoStop}%
\bibitem [{\citenamefont {Huang}\ \emph {et~al.}(2017)\citenamefont {Huang},
  \citenamefont {Clark}, \citenamefont {Navarro-Moratalla}, \citenamefont
  {Klein}, \citenamefont {Cheng}, \citenamefont {Seyler}, \citenamefont
  {Zhong}, \citenamefont {Schmidgall}, \citenamefont {McGuire}, \citenamefont
  {Cobden}, \citenamefont {Yao}, \citenamefont {Xiao}, \citenamefont
  {Jarillo-Herrero},\ and\ \citenamefont {Xu}}]{Huang2017}%
  \BibitemOpen
  \bibfield  {author} {\bibinfo {author} {\bibfnamefont {B.}~\bibnamefont
  {Huang}}, \bibinfo {author} {\bibfnamefont {G.}~\bibnamefont {Clark}},
  \bibinfo {author} {\bibfnamefont {E.}~\bibnamefont {Navarro-Moratalla}},
  \bibinfo {author} {\bibfnamefont {D.~R.}\ \bibnamefont {Klein}}, \bibinfo
  {author} {\bibfnamefont {R.}~\bibnamefont {Cheng}}, \bibinfo {author}
  {\bibfnamefont {K.~L.}\ \bibnamefont {Seyler}}, \bibinfo {author}
  {\bibfnamefont {D.}~\bibnamefont {Zhong}}, \bibinfo {author} {\bibfnamefont
  {E.}~\bibnamefont {Schmidgall}}, \bibinfo {author} {\bibfnamefont {M.~A.}\
  \bibnamefont {McGuire}}, \bibinfo {author} {\bibfnamefont {D.~H.}\
  \bibnamefont {Cobden}}, \bibinfo {author} {\bibfnamefont {W.}~\bibnamefont
  {Yao}}, \bibinfo {author} {\bibfnamefont {D.}~\bibnamefont {Xiao}}, \bibinfo
  {author} {\bibfnamefont {P.}~\bibnamefont {Jarillo-Herrero}}, \ and\ \bibinfo
  {author} {\bibfnamefont {X.}~\bibnamefont {Xu}},\ }\href {\doibase
  10.1038/nature22391} {\bibfield  {journal} {\bibinfo  {journal} {Nature}\
  }\textbf {\bibinfo {volume} {546}},\ \bibinfo {pages} {270} (\bibinfo {year}
  {2017})}\BibitemShut {NoStop}%
\bibitem [{\citenamefont {Davis}\ and\ \citenamefont
  {Narath}(1964)}]{PhysRev.134.A433}%
  \BibitemOpen
  \bibfield  {author} {\bibinfo {author} {\bibfnamefont {H.~L.}\ \bibnamefont
  {Davis}}\ and\ \bibinfo {author} {\bibfnamefont {A.}~\bibnamefont {Narath}},\
  }\href {\doibase 10.1103/PhysRev.134.A433} {\bibfield  {journal} {\bibinfo
  {journal} {Phys. Rev.}\ }\textbf {\bibinfo {volume} {134}},\ \bibinfo {pages}
  {A433} (\bibinfo {year} {1964})}\BibitemShut {NoStop}%
\bibitem [{\citenamefont {Narath}\ and\ \citenamefont
  {Davis}(1965)}]{PhysRev.137.A163}%
  \BibitemOpen
  \bibfield  {author} {\bibinfo {author} {\bibfnamefont {A.}~\bibnamefont
  {Narath}}\ and\ \bibinfo {author} {\bibfnamefont {H.~L.}\ \bibnamefont
  {Davis}},\ }\href {\doibase 10.1103/PhysRev.137.A163} {\bibfield  {journal}
  {\bibinfo  {journal} {Phys. Rev.}\ }\textbf {\bibinfo {volume} {137}},\
  \bibinfo {pages} {A163} (\bibinfo {year} {1965})}\BibitemShut {NoStop}%
\bibitem [{\citenamefont {Samuelsen}\ \emph {et~al.}(1971)\citenamefont
  {Samuelsen}, \citenamefont {Silberglitt}, \citenamefont {Shirane},\ and\
  \citenamefont {Remeika}}]{PhysRevB.3.157}%
  \BibitemOpen
  \bibfield  {author} {\bibinfo {author} {\bibfnamefont {E.~J.}\ \bibnamefont
  {Samuelsen}}, \bibinfo {author} {\bibfnamefont {R.}~\bibnamefont
  {Silberglitt}}, \bibinfo {author} {\bibfnamefont {G.}~\bibnamefont
  {Shirane}}, \ and\ \bibinfo {author} {\bibfnamefont {J.~P.}\ \bibnamefont
  {Remeika}},\ }\href {\doibase 10.1103/PhysRevB.3.157} {\bibfield  {journal}
  {\bibinfo  {journal} {Phys. Rev. B}\ }\textbf {\bibinfo {volume} {3}},\
  \bibinfo {pages} {157} (\bibinfo {year} {1971})}\BibitemShut {NoStop}%
\bibitem [{\citenamefont {Chen}\ \emph {et~al.}(2020)\citenamefont {Chen},
  \citenamefont {Chung}, \citenamefont {Chen}, \citenamefont {Duan},
  \citenamefont {Schneidewind}, \citenamefont {Radelytskyi}, \citenamefont
  {Voneshen}, \citenamefont {Ewings}, \citenamefont {Stone}, \citenamefont
  {Kolesnikov}, \citenamefont {Winn}, \citenamefont {Chi}, \citenamefont
  {Mole}, \citenamefont {Yu}, \citenamefont {Gao},\ and\ \citenamefont
  {Dai}}]{PhysRevB.101.134418}%
  \BibitemOpen
  \bibfield  {author} {\bibinfo {author} {\bibfnamefont {L.}~\bibnamefont
  {Chen}}, \bibinfo {author} {\bibfnamefont {J.-H.}\ \bibnamefont {Chung}},
  \bibinfo {author} {\bibfnamefont {T.}~\bibnamefont {Chen}}, \bibinfo {author}
  {\bibfnamefont {C.}~\bibnamefont {Duan}}, \bibinfo {author} {\bibfnamefont
  {A.}~\bibnamefont {Schneidewind}}, \bibinfo {author} {\bibfnamefont
  {I.}~\bibnamefont {Radelytskyi}}, \bibinfo {author} {\bibfnamefont {D.~J.}\
  \bibnamefont {Voneshen}}, \bibinfo {author} {\bibfnamefont {R.~A.}\
  \bibnamefont {Ewings}}, \bibinfo {author} {\bibfnamefont {M.~B.}\
  \bibnamefont {Stone}}, \bibinfo {author} {\bibfnamefont {A.~I.}\ \bibnamefont
  {Kolesnikov}}, \bibinfo {author} {\bibfnamefont {B.}~\bibnamefont {Winn}},
  \bibinfo {author} {\bibfnamefont {S.}~\bibnamefont {Chi}}, \bibinfo {author}
  {\bibfnamefont {R.~A.}\ \bibnamefont {Mole}}, \bibinfo {author}
  {\bibfnamefont {D.~H.}\ \bibnamefont {Yu}}, \bibinfo {author} {\bibfnamefont
  {B.}~\bibnamefont {Gao}}, \ and\ \bibinfo {author} {\bibfnamefont
  {P.}~\bibnamefont {Dai}},\ }\href {\doibase 10.1103/PhysRevB.101.134418}
  {\bibfield  {journal} {\bibinfo  {journal} {Phys. Rev. B}\ }\textbf {\bibinfo
  {volume} {101}},\ \bibinfo {pages} {134418} (\bibinfo {year}
  {2020})}\BibitemShut {NoStop}%
\bibitem [{\citenamefont {Owerre}(2017)}]{Owerre2017}%
  \BibitemOpen
  \bibfield  {author} {\bibinfo {author} {\bibfnamefont {S.~A.}\ \bibnamefont
  {Owerre}},\ }\href {\doibase 10.1088/2399-6528/aa8843} {\bibfield  {journal}
  {\bibinfo  {journal} {Journal of Physics Communications}\ }\textbf {\bibinfo
  {volume} {1}},\ \bibinfo {pages} {021002} (\bibinfo {year} {2017})},\
  \bibinfo {note} {arXiv:1705.04694 [cond-mat]}\BibitemShut {NoStop}%
\bibitem [{\citenamefont {Koretsune}\ \emph {et~al.}(2015)\citenamefont
  {Koretsune}, \citenamefont {Nagaosa},\ and\ \citenamefont
  {Arita}}]{Koretsune2015}%
  \BibitemOpen
  \bibfield  {author} {\bibinfo {author} {\bibfnamefont {T.}~\bibnamefont
  {Koretsune}}, \bibinfo {author} {\bibfnamefont {N.}~\bibnamefont {Nagaosa}},
  \ and\ \bibinfo {author} {\bibfnamefont {R.}~\bibnamefont {Arita}},\ }\href
  {\doibase 10.1038/srep13302} {\bibfield  {journal} {\bibinfo  {journal}
  {Scientific Reports}\ }\textbf {\bibinfo {volume} {5}},\ \bibinfo {pages}
  {13302} (\bibinfo {year} {2015})}\BibitemShut {NoStop}%
\bibitem [{\citenamefont {Legrand}\ \emph {et~al.}(2022)\citenamefont
  {Legrand}, \citenamefont {Sassi}, \citenamefont {Ajejas}, \citenamefont
  {Collin}, \citenamefont {Bocher}, \citenamefont {Jia}, \citenamefont
  {Hoffmann}, \citenamefont {Zimmermann}, \citenamefont {Blügel},
  \citenamefont {Reyren}, \citenamefont {Cros},\ and\ \citenamefont
  {Thiaville}}]{Legrand2022}%
  \BibitemOpen
  \bibfield  {author} {\bibinfo {author} {\bibfnamefont {W.}~\bibnamefont
  {Legrand}}, \bibinfo {author} {\bibfnamefont {Y.}~\bibnamefont {Sassi}},
  \bibinfo {author} {\bibfnamefont {F.}~\bibnamefont {Ajejas}}, \bibinfo
  {author} {\bibfnamefont {S.}~\bibnamefont {Collin}}, \bibinfo {author}
  {\bibfnamefont {L.}~\bibnamefont {Bocher}}, \bibinfo {author} {\bibfnamefont
  {H.}~\bibnamefont {Jia}}, \bibinfo {author} {\bibfnamefont {M.}~\bibnamefont
  {Hoffmann}}, \bibinfo {author} {\bibfnamefont {B.}~\bibnamefont
  {Zimmermann}}, \bibinfo {author} {\bibfnamefont {S.}~\bibnamefont {Blügel}},
  \bibinfo {author} {\bibfnamefont {N.}~\bibnamefont {Reyren}}, \bibinfo
  {author} {\bibfnamefont {V.}~\bibnamefont {Cros}}, \ and\ \bibinfo {author}
  {\bibfnamefont {A.}~\bibnamefont {Thiaville}},\ }\href {\doibase
  10.1103/physrevmaterials.6.024408} {\bibfield  {journal} {\bibinfo  {journal}
  {Physical Review Materials}\ }\textbf {\bibinfo {volume} {6}},\ \bibinfo
  {pages} {024408} (\bibinfo {year} {2022})}\BibitemShut {NoStop}%
\bibitem [{\citenamefont {Ant{\~{a}}o}\ and\ \citenamefont
  {Peres}(2023)}]{Antao2023}%
  \BibitemOpen
  \bibfield  {author} {\bibinfo {author} {\bibfnamefont {T.~V.~C.}\
  \bibnamefont {Ant{\~{a}}o}}\ and\ \bibinfo {author} {\bibfnamefont
  {N.~M.~R.}\ \bibnamefont {Peres}},\ }\href {\doibase
  10.1103/PhysRevB.107.235410} {\bibfield  {journal} {\bibinfo  {journal}
  {Physical Review B}\ }\textbf {\bibinfo {volume} {107}},\ \bibinfo {pages}
  {235410} (\bibinfo {year} {2023})},\ \Eprint
  {http://arxiv.org/abs/2303.03305} {arXiv:2303.03305} \BibitemShut {NoStop}%
\bibitem [{\citenamefont {Zhuo}\ \emph {et~al.}(2021)\citenamefont {Zhuo},
  \citenamefont {Li},\ and\ \citenamefont {Manchon}}]{Zhuo2021}%
  \BibitemOpen
  \bibfield  {author} {\bibinfo {author} {\bibfnamefont {F.}~\bibnamefont
  {Zhuo}}, \bibinfo {author} {\bibfnamefont {H.}~\bibnamefont {Li}}, \ and\
  \bibinfo {author} {\bibfnamefont {A.}~\bibnamefont {Manchon}},\ }\href
  {\doibase 10.1103/PhysRevB.104.144422} {\bibfield  {journal} {\bibinfo
  {journal} {Physical Review B}\ }\textbf {\bibinfo {volume} {104}},\ \bibinfo
  {pages} {144422} (\bibinfo {year} {2021})}\BibitemShut {NoStop}%
\bibitem [{\citenamefont {Zhuo}\ \emph {et~al.}(2022)\citenamefont {Zhuo},
  \citenamefont {Li},\ and\ \citenamefont {Manchon}}]{Zhuo2022}%
  \BibitemOpen
  \bibfield  {author} {\bibinfo {author} {\bibfnamefont {F.}~\bibnamefont
  {Zhuo}}, \bibinfo {author} {\bibfnamefont {H.}~\bibnamefont {Li}}, \ and\
  \bibinfo {author} {\bibfnamefont {A.}~\bibnamefont {Manchon}},\ }\href
  {\doibase 10.1088/1367-2630/ac51a8} {\bibfield  {journal} {\bibinfo
  {journal} {New Journal of Physics}\ }\textbf {\bibinfo {volume} {24}},\
  \bibinfo {pages} {023033} (\bibinfo {year} {2022})}\BibitemShut {NoStop}%
\end{thebibliography}%

\end{document}